\documentclass[preprint]{aastex}
\usepackage{graphicx}   
\usepackage{verbatim}
\graphicspath{{Figures/}}
\usepackage{bm}

\newcommand{\dat}{{\bf x}}
\newcommand{\simdat}{{\bf x^*}}
\newcommand{\summ}{{s}}
\newcommand{\simsumm}{{s^*}}
\newcommand{\param}{\theta}
\newcommand{\simparam}{\theta^*}
\newcommand{\cond}{\hspace{.01in}|\:}
\newcommand{\ztru}{\zeta}
\newcommand{\zobs}{z}
\newcommand{\zspec}{z^{\rm spec}}
\newcommand{\zphot}{z^{\rm phot}}

\newcommand{\bftau}{\boldsymbol{\tau}}
\newcommand{\sigmdm}{\left ( \sigma^{z}_{\mu,i} \right )^2}
\newcommand{\dens}{p} 
 
\begin{document}  

\title{Likelihood-Free Cosmological Inference with Type Ia Supernovae: Approximate Bayesian Computation for a Complete Treatment of Uncertainty}
\author{Anja Weyant\altaffilmark{1}, Chad Schafer\altaffilmark{2}, W. Michael Wood-Vasey\altaffilmark{1}}
   
\affil{1 Pittsburgh Particle Physics, Astrophysics, and Cosmology Center (PITT PACC),  Physics and Astronomy Department, University of Pittsburgh, Pittsburgh, PA 15260, USA}
\affil{2 Department of Statistics, Carnegie Mellon University, Pittsburgh, PA 15213, USA}

\email{anw19@pitt.edu}

\keywords{supernova,cosmology,techniques}

\begin{abstract}
Cosmological inference becomes increasingly difficult when complex data-generating processes cannot be
modeled by simple probability distributions.
With the ever-increasing size of data sets in cosmology, there is increasing burden placed on 
adequate modeling; systematic errors in the model will dominate where previously these 
were swamped by statistical errors.
For example, Gaussian distributions are an insufficient representation for errors in 
quantities like photometric redshifts. Likewise, it can be difficult to quantify analytically
the distribution of errors that are introduced 
in complex fitting codes.
Without a simple form for these distributions, it
becomes difficult to accurately construct a likelihood function for the data as a function
of parameters of interest.
Approximate Bayesian computation (ABC) provides a means of probing the posterior 
distribution when direct calculation of a sufficiently accurate likelihood is intractable.
ABC allows one to bypass direct calculation of the likelihood but instead relies upon
the ability to simulate the forward process that generated the data.
These simulations can naturally incorporate priors placed on nuisance parameters, and
hence these can be marginalized in a natural way.
We present and discuss ABC methods in the context of supernova cosmology using data from the SDSS-II Supernova Survey.  
Assuming a flat cosmology and constant dark energy equation of state we demonstrate that ABC can recover an accurate posterior distribution.  Finally we show that ABC can still produce an accurate posterior distribution when we contaminate the sample with Type IIP supernovae. 
\end{abstract}

\section{Introduction} 

Since the discovery of the accelerated expansion of our universe \citep{Riess98, Perlmutter99}, 
the quality of Type Ia supernova (SN~Ia) data sets has improved and the quantity has grown to thousands through individual efforts with 
the Hubble Space Telescope \citep{Knop03, Riess04, Amanullah10} 
and surveys such as
the Supernova Legacy Survey \citep{Astier06, Conley11}, 
the ESSENCE Supernova Survey \citep{Miknaitis07,Wood-Vasey07}, 
the CfA Supernova group \citep{Hicken09b, Hicken12}, 
the Carnegie Supernova Project \citep{Contreras10,Stritzinger11}, 
the Sloan Digital Sky Survey-II \citep{Lampeitl10}, 
and the Lick Observatory Supernova Search \citep{Gane10}.
Additional current and near-future surveys such as 
the Palomar Transient Factory\footnote{\url{http://www.astro.caltech.edu/ptf/}}~\citep{law09},
the Panoramic Survey Telescope and Rapid Response System (Pan-STARRS)\footnote{\url{http://pan-starrs.ifa.hawaii.edu/public/}},
SkyMapper\footnote{\url{http://www.mso.anu.edu.au/skymapper/}}, 
and the Dark Energy Survey\footnote{\url{http://www.darkenergysurvey.org/}}
will increase the sample by another order of magnitude with the goal of obtaining tighter constraints on the nature of dark energy.
The Large Synoptic Survey Telescope (LSST) anticipates 
observing hundreds of thousands of well-measured Type Ia supernovae (SNe~Ia) \citep{LSSTscienceBook}.  

In this new regime of large numbers of SNe~Ia the weaknesses and limitations of our current $\chi^2$ likelihood approach to estimating cosmological parameters are becoming apparent.  For example, with limited spectroscopic follow-up, we must rely on light-curve classification codes and photometric redshift tools to maximize the scientific potential of SN~Ia cosmology with LSST and near-future surveys.  These two crucial steps alone introduce a nontrivial component to our probability models from which we construct the likelihood.  Additionally, there are significant systematic uncertainties including errors from calibration, survey design and cadence, host galaxy subtraction and intrinsic dust, population evolution, gravitational lensing, and peculiar velocities.  All of these uncertainties contribute to a probability model which simply cannot be accurately described by a multivariate normal distribution.    

In this paper we describe how the statistical technique of Approximate Bayesian Computation (ABC) can be used to overcome these challenges and explore the space of cosmological parameters in the face of non-Gaussian distributions of systematic uncertainties, complicated functional priors, and large data sets.  We encourage the reader to read the recent paper by \citet{Cameron12} for an introduction to and application of ABC in the context of galaxy evolution.  We here focus on supernova cosmology, but ABC has applicability in a wide range of forward-modeling problems in astrophysics and cosmology.  


\subsection{Classical Estimation of Cosmological Parameters from SN~Ia Data}
\label{sec:ClassicalSNIaCosmology}

Cosmological inference with SNe~Ia is a classical statistical estimation problem.
We have data, our set of supernova light-curve observations, and we seek to infer 
something about the universe in which we live. 
It is standard in cosmology to adopt a Bayesian approach to inference. 
To clarify our basic conceptual and notational framework, we review Bayes theorem, 
a simple identity which relates the posterior probability distribution--the probability of a set of model parameters given the data--to the probability of the data given the model, the likelihood.  
More precisely, the posterior probability distribution is derived as
\begin{equation} \pi({\theta}\cond \dat) = \frac{\dens(\dat \cond {\theta}) 
   \pi({\theta})}{\dens(\dat)} \end{equation}
where $\dens(\dat \cond \theta)$ is the likelihood, $\pi({\theta})$ is the prior 
on the vector of model parameters $\theta$, 
and $\dens(\dat)$ is the marginal 
probability of the data $\dat$ ($\dens(\dat) = \int_{{\Theta}}\dens(\dat \cond \theta)\pi(\theta)d{\theta}$).  
The Bayesian framework is powerful in that it allows evidence and experience to modify the prior.
The approach is challenging, however, in that standard computation methods rely upon full specification
of the likelihood $\dens(\dat \cond \theta)$; this can be challenging in applications of interest.

For example, consider a cosmological model for which the distance modulus can be written as $\mu_{\rm model} = \mu_{\rm model}(\Omega_M, \Omega_{\Lambda}, w, z)$.  If we assume that each measured $\mu$ has a probability distribution function (PDF) described by a Gaussian with standard deviation $\sigma$ we can write the likelihood for a single observation as 
\begin{equation}
p(\mu_i, z_i | \Omega_M, \Omega_{\Lambda}, w) \propto \exp{\left [-\frac{(\mu_i - \mu_{\rm model}(z_i, \Omega_M, \Omega_{\Lambda}, w))^2}{2\sigma_i^2} \right ]}.
\end{equation}
If the distance observations are independent after calibration such that there are no correlated uncertainties 
we can simply multiply the likelihood of each observation together.  By taking the logarithm, we can write a more convenient form of the likelihood as follows 
\begin{equation} -2\ln{(p(\mu, z | \Omega_M, \Omega_{\Lambda}, w) )} = K + \sum_{i=1}^N \frac{(\mu_i - \mu_{\rm model}(z_i, \Omega_M, \Omega_{\Lambda}, w))^2}{\sigma_i^2}, \label{eq:chisq} \end{equation}
where $K$ is an unimportant constant, giving us the familiar $\chi^2$ statistic.  
Note that the use of this form of the likelihood function and $\chi^2$ statistic is based on the assumption 
of independent data with normally distributed uncertainties.

Traditionally when making cosmological inference with SNe~Ia one calculates the $\chi^2$ statistic \citep{Conley11, Kessler09b, Wood-Vasey07, Astier06, Riess04}.
One method of including systematic uncertainties in such a framework is to use the ``quadrature'' method, accurately named by \citet{Conley11}. Systematic errors which are not redshift dependent and add scatter to the overall Hubble diagram are added in quadrature to the statistical uncertainties.  For other sources of systematic uncertainty it is typical to perform the analysis with and without including the systematic effect on the data.  The difference in inferred cosmological parameter is then a measure of the systematic uncertainty.  All systematic effects are then added in quadrature as the quoted total systematic uncertainty.  This method has been used in recent cosmological analyses by \citet{Kessler09b}, \citet{Wood-Vasey07}, and \citet{Astier06}.  It has the advantage of being simple to implement but the disadvantage of missing correlations between systematic uncertainties, not producing the full likelihood, and could be inappropriate for asymmetric error distributions \citep{barlow03}.
One also has the difficult task of estimating the size of the systematic uncertainty and implementing its effect in the analysis.

\citet{Conley11} presented a more thorough approach to incorporating systematic uncertainties into a $\chi^2$ analysis using a covariance matrix.  By implementing a covariance matrix one can drop the assumption of independent data in Eq. \ref{eq:chisq}.  The covariance matrix can be decomposed into a diagonal, statistical component and two off-diagonal matrices which include statistical and systematic uncertainty.  These off-diagonal covariance matrices include uncertainties from, e.g., uncertainty in the supernova model which is statistical in nature but could be correlated between different SNe~Ia and uncertainty in zero points which would systematically affect all SNe~Ia.
\citet{Kowalski08} and \citet{Amanullah10} present similar methods which are approximations to \citet{Conley11}'s covariance matrix approach.  However, the overall approach must be modified for uncertainties due to, e.g., type contamination and Malmquist bias.  They have the effect of adding or removing supernovae from the sample which is difficult to represent in a covariance matrix.
For systematic effects such as these the field of supernova cosmology is moving toward calculating the corrections to the data using artificial SNe~Ia generated from Monte Carlo simulations.

Bayesian inference becomes increasingly difficult as we depart from normal error distributions or when the likelihood function is not analytically or computationally tractable.   Direct calculation of the likelihood may involve many integrations over systematic uncertainties, nuisance parameters, and latent variables.  These integrations can make the use of standard Markov Chain Monte Carlo (MCMC) techniques very challenging and computational expensive.
It may also be incredibly difficult to construct an analytic probability model over which to marginalize.

ABC allows one to bypass direct calculation of the likelihood by simulating data from the posterior distribution.  The posterior distribution is then constructed from the model parameters necessary to simulate data which resemble the observed data. By incorporating into the simulation all of the statistical and systematic uncertainties for which we have models and priors, the simulation {\it knows} about the complicated probability model even thought the observer may not be able to have the model written out as a set of equations or numerical integrals.  By simulating many realistic datasets one can marginalize over the nuisance parameters and systematic uncertainties such that high-dimensional marginalization problems, as in population genetics for which ABC techniques were first developed, are now computationally feasible.   ABC is a consistent framework to incorporate systematic uncertainties with the cosmological model and more clearly defines what it means to use Monte Carlo simulations of artificial SNe~Ia to quantify systematic uncertainty.

We begin in \S~\ref{sec:SNeToy} by motivating the general problem and discussing the breakdown of current cosmological inference methods using a simple example.  In \S~\ref{sec:ABCmethod} we outline three separate ABC algorithms and discuss their merits.  To provide the reader with an introductory example of using ABC, we then illustrate how one might perform cosmological inference with Sequential Monte Carlo (SMC) ABC using the simple model discussed in \S~\ref{sec:SNeToy}.  In \S~\ref{sec:ABCwSDSS} we present a more sophisticated analysis using SNe~Ia from the Sloan Digital Sky Survey-II (SDSS-II) Supernova Survey and demonstrate how one might perform cosmological inference with a tool like the SuperNova ANAlysis (SNANA) \citep{Kessler09a} software using SMC ABC techniques.  We compare our results to the cosmological analysis performed in \citet{Kessler09b} using statistical errors only.  At the end of this section we show that ABC can recover the full posterior distribution when we contaminate the data with simulated Type IIP supernovae.  We discuss directions for future work in \S~\ref{sec:future_work} and conclude in \S~\ref{sec:Conclusion}.



\section{General Problem Formulation}
\label{sec:SNeToy}

Here we establish notation that we will use in discussing the SN~Ia inference
problem. Below we explain how this framework could be extended to other cosmological
inference challenges.
Let $\mu_i$ be the measured distance modulus of the $i^{th}$ 
SN~Ia in our sample, $\tau_i$ be its true distance modulus, $\zobs_i$ be 
the estimated redshift, and $\theta$ be the vector of cosmological parameters.
We will use bold faced variables to indicate a set of $n$ supernovae, e.g., ${\bf\zobs}= \{ \zobs_1,...,\zobs_n  \}$.
Here, we stress that the ``estimated redshift'' will be, in practice, the 
redshift as estimated from photometry, i.e., the photometric redshift.

The underlying objective is to determine the posterior of the cosmological
parameters $\theta$ given the
observed data $({\boldsymbol{\mu}}, {\bf \zobs})$.
There are two natural analytical routes, both of which lead to the same challenges.
The first route is to note that
the posterior of $\theta$ can be decomposed as
\begin{equation}
   \pi (\theta \cond {\boldsymbol{\mu}}, {\bf \zobs}) = K \dens ({\boldsymbol{\mu}} \cond \theta, {\bf \zobs}) 
\pi (\theta, {\bf \zobs})
\end{equation}
where $K$ is a constant that does not depend on $\theta$ and
\begin{eqnarray}
   \dens ({\boldsymbol{\mu}} \cond \theta, {\bf \zobs})  \pi(\theta, {\bf \zobs})  
   & = & \left[ \int \dens ({\boldsymbol{\mu}} \cond \theta, {\bf \zobs}, \bftau) \:\dens({\bftau} \cond \theta, {\bf \zobs})\: d{\bftau}\right]
\pi (\theta, {\bf \zobs}) \\
   & = & \left[ \int \dens({\boldsymbol{\mu}} \cond \theta, {\bftau}) \:\dens({\bftau} \cond \theta, {\bf \zobs})\: d{\bftau}\right]
\pi (\theta, {\bf \zobs}).
 \label{eq:postdec}
\end{eqnarray}
Note that in this last step, the density of ${\boldsymbol{\mu}}$ conditional on $\theta$ and ${\bf \zobs}$ is replaced with
the density of ${\boldsymbol{\mu}}$ conditional only on $\theta$. Here we are assuming
that ${\boldsymbol{\mu}}$ and ${\bf \zobs}$ are independent given ${\bftau}$: Once $\bftau$ is known, the information
in ${\bf \zobs}$ does not affect the distribution of ${\boldsymbol{\mu}}$.  We note that this assumption is not true if one is using the photometric redshift determined from the supernova light curve.

We could pose this problem in general statistical terms as follows.
Assume that ${\boldsymbol{\mu}} = \left \{ \mu_1, \mu_2, \ldots, \mu_n \right \}$ 
are random variables such that
the distribution of $\mu_i$ is determined by parameters $\theta$ and $\tau_i$.
Here, $\theta$ represents the unknown parameters 
common to the $\mu_i$ while $\tau_i$ are the object-specific parameters. 
We further assume the existence of
additional data, denoted $\zobs_i$, which have
the property that $\mu_i$ and $\zobs_i$ are independent conditional on $\tau_i$.
The quantities $\zobs_i$ can be thought of as properties that help in the 
estimation of $\tau_i$, but would not be useful for estimating $\theta$ if
$\tau_i$ were known. 

Note that each of $\mu_i$ and $\tau_i$ could be vectors.
For example, in \citet{Mandel11}, $\mu_i$ stores the full observed light curve of
the supernova and $\tau_i$ comprises not only the true distance modulus, but also parameters
that capture the effect of extinction and dust and 
that define the true, underlying light curve.
As mentioned above, these have the property that, if $\tau_i$ were known, $\zobs_i$
would not provide useful additional information for the estimation of $\theta$.



The second route is to rewrite the posterior as 
\begin{equation}
   \pi(\theta \cond {\boldsymbol{\mu}}, {\bf \zobs}) = \int \dens(\theta, {\bftau} \cond {\boldsymbol{\mu}}, {\bf \zobs}) \:d{\bftau} 
\end{equation}
and then rely upon the fact that, as derived above,
\begin{equation}
   \dens(\theta, {\bftau} \cond {\boldsymbol{\mu}}, {\bf \zobs}) = 
   \dens({\boldsymbol{\mu}} \cond \theta, {\bftau}) \:\dens({\bftau} \cond \theta, {\bf \zobs})
   \pi(\theta, {\bf \zobs})
\end{equation}
to construct a hierarchical Bayesian model for the unknown ``parameters'' which now consist
of both $\theta$ and $\bftau$. 
To analytically obtain the posterior in terms of only $\theta$, one
must integrate over $\bftau$, i.e., find
\begin{equation}
   \int \dens ({\boldsymbol{\mu}} \cond \theta, {\bftau}) \:\dens({\bftau} \cond \theta, {\bf \zobs})\: d{\bftau}.
   \label{eq:margint}
\end{equation}
This is exactly the form of the challenging integral that was confronted above in Equation (\ref{eq:postdec}).
One can often justify further conditional independence assumptions and write
\begin{eqnarray}
   \int \dens({\boldsymbol{\mu}} \cond \theta, {\bftau}) \:\dens({\bftau} \cond \theta, {\bf \zobs})\: d{\bftau}\:
   & = & 
   \int \prod_{i=1}^n \dens(\mu_i \cond \theta, \tau_i) \:\dens(\tau_i \cond \theta, \zobs_i)\: d{\bftau} \\
   & = &
   \:\prod_{i=1}^n \int \dens(\mu_i \cond \theta, \tau_i) \:\dens(\tau_i \cond \theta, \zobs_i)\: d\tau_i.
\end{eqnarray}
Still, the computational feasibility of using analytical approaches to finding the posterior for $\theta$ will
depend on the form of
\begin{equation}
\dens(\mu_i \cond \theta, \zobs_i) =  \int \dens(\mu_i \cond \theta, \tau_i) \:\dens(\tau_i \cond \theta, \zobs_i)\: d\tau_i.\\
   \label{intoverz}
\end{equation}
In practice, the complex nature of photometric redshift estimators will yield a
complex form for the distribution
$\dens(\tau_i \cond \theta, \zobs_i)$.

An alternative is to adopt the ``second route'' described above but instead
utilize MCMC methods to simulate from the posterior for both $(\theta, \tau)$.
This is the approach taken in \citet{Mandel11}.
This avoids the integral over $\tau_i$, but it is still apparent that
practical implementation of analytical or MCMC methods when $n$ is large (and hence $\bftau$ is of 
high dimension)
forces one to make choices for $\dens(\mu_i \cond \theta, \tau_i)$ and $\dens(\tau_i \cond \theta, \zobs_i)$
which may not be realistic. 
Unfortunately, as $n$ gets large, even small mistakes in the specification of these
densities could lead to significant biases in the estimates of the parameters. This
is one of the fundamental challenges facing cosmology as we are presented with ever-larger
data sets. In what follows we will develop an example that illustrates this point.

\subsection{A Simple Example}

To begin, note that in the present example $\mu_i$ is the measured distance modulus, $\zobs_i$ 
is the measured redshift, $\tau_i$ is the true distance modulus, and $\theta$ represent 
the set of cosmological parameters.  We ignore for the moment all parameters which affect 
the measured distance modulus except $\zobs_i$ and $\theta$.
The measured redshift $\zobs_i$
may differ from the true redshift of the supernova, which we will denote $\ztru_i$.
Consider the following three scenarios: 
\begin{enumerate}
\item {\it $\zobs_i = \ztru_i$, i.e., the redshift is known exactly.} In this case, and
under our simplifying assumptions, we know exactly the value of $\tau_i$, and hence the ``density''
$\dens(\tau_i \cond \theta, \zobs_i)$ is a delta function at this known value.
\item The redshift is observed with some normal error.  We model $\ztru_i$ with a Gaussian PDF with mean $\zobs_i$ and variance $\sigma_{z,i}^2$. In this case we can apply the
so-called ``delta method'' and state that $\dens(\tau_i \cond \theta, \zobs_i)$ is approximately Gaussian
with mean $\mu(\zobs_i, \theta)$.  This scenario is analogous to measuring a spectroscopic redshift with a small error such that a Gaussian approximation for the PDF of $\ztru_i$ is sufficient or a photometric redshift which has a PDF which can be modeled well by a Gaussian.
\item $\zobs_i$ is observed with some complicated uncertainty. The PDF is not described by a simple function although $\dens(\tau_i \cond \theta, \zobs_i)$ may be estimated using observed data.  This is the case for most photometric redshifts.
\end{enumerate}
Of course, the first case is unrealistic.
In order to demonstrate the pitfalls of making unwarranted
assumptions regarding the likelihood function, we will first focus on the second case, in particular assume that
  $\dens(\tau_i \cond \theta, \zobs_i)$
is a Gaussian density with mean $\mu(\zobs_i, \theta)$.
The rationale for this approximation relies on the 
assumption that the true redshift $\ztru_i$ also has a Gaussian distribution,
in this case with mean $\zobs_i$ and variance $\sigma^2_{z,i}$.  
The true distance modulus is $\tau_i = \mu(\ztru_i, \theta)$, so, using the standard linear approximation,
we can argue that $\tau_i$ is approximately normal with mean $\mu(\zobs_i,\theta)$ and variance
\begin{equation}
   \sigmdm =  \left[ \frac{\partial \mu(\zobs_i, \theta)}{\partial \zobs_i}\right]^2 \sigma^2_{z,i}.
\end{equation}
Then, the observed distance modulus can be modeled as the true distance modulus plus some
additional Gaussian error; this is taken to have mean zero and
variance $\left ( \sigma_{\mu,i} \right )^2$. In a real-life application this variance includes uncertainty from the observed intrinsic dispersion in distance modulus and uncertainty from fitting the light curve. 

This is the current approach in most cosmological analyses where one has spectroscopic redshifts 
for each SN~Ia \citep{Conley11, Kessler09b, Wood-Vasey07, Astier06}.  The uncertainty in redshift 
is transferred to the uncertainty in measured distance modulus and one can find an analytic 
solution to Eq.~\ref{intoverz} by noting that the integral is simply the convolution of two
normal densities. Hence the result of Eq.~\ref{intoverz} is another normal density, but now
with mean $\mu(z_i, \theta)$ and variance $\sigmdm + \left ( \sigma_{\mu,i} \right )^2$.
This approach is also possible for larger uncertainties like those from photometric redshifts,
but the concern becomes the fact that the linear approximation utilized does not extend to
larger ranges of redshift. In what follows 
we examine the consequences of making this Gaussian assumption for photometric redshift 
uncertainties when the approximation is not valid, i.e., we treat scenario 3 as if it were scenario 2.

\begin{figure}
\plotone{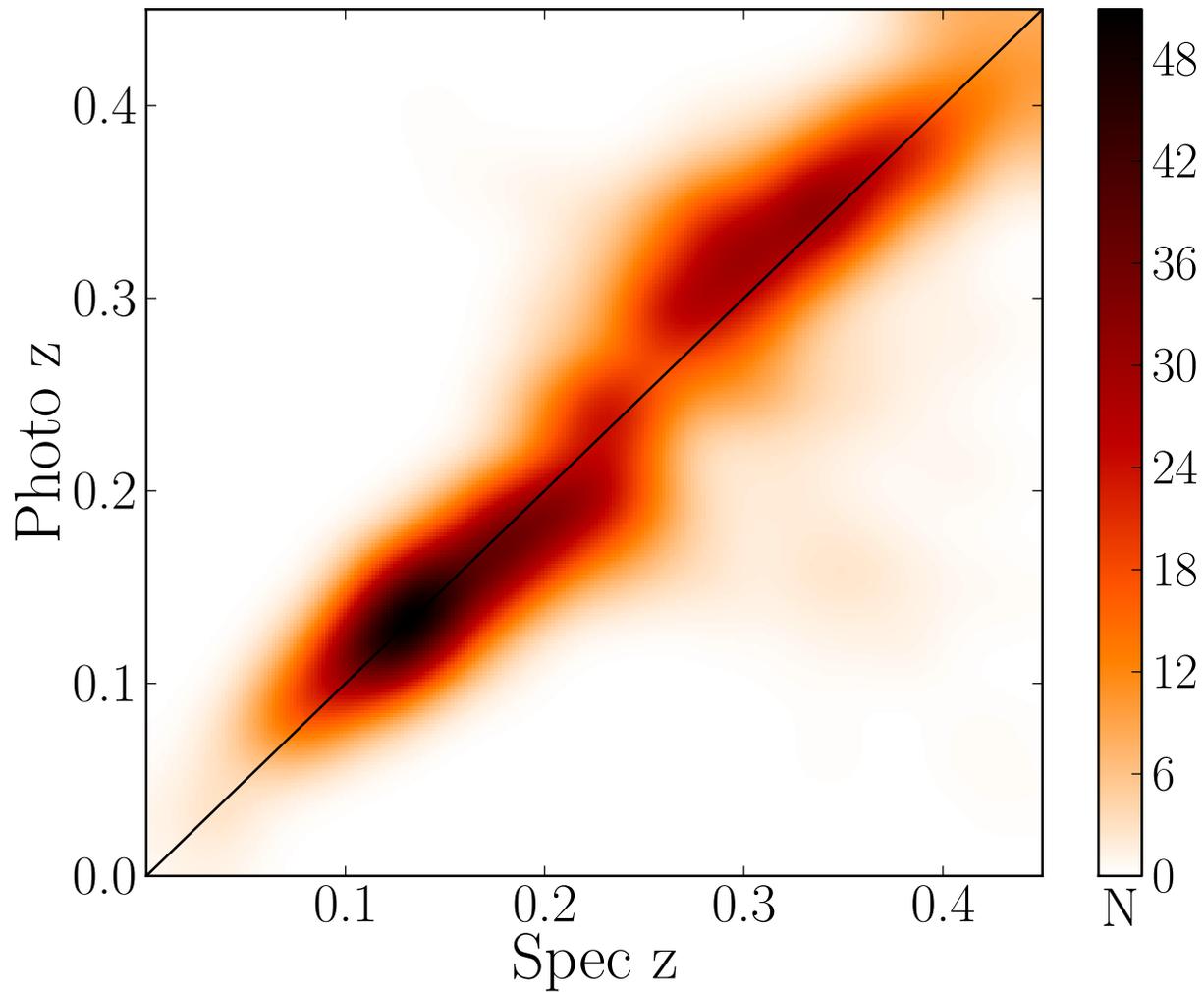}
\caption{Photometric vs. spectroscopic redshift for 1744 simulated SNe~Ia using SNANA and smoothed with a Gaussian kernel.  Note the complex structure and asymmetry about the one-to-one line indicating departures from Gaussianity.  This sample is used to represent a realistic joint distribution between the spectroscopic and photometric redshifts.  }
\label{fig:speczphotz}
\end{figure}

Fig.~\ref{fig:speczphotz} shows the photometric versus spectroscopic redshift for a sample of 1744 SNe~Ia generated using SNANA\footnote{\url{http://sdssdp62.fnal.gov/sdsssn/SNANA-PUBLIC/}} version v9\_32 and smoothed with a Gaussian kernel. To make this figure, light curves were simulated and fit from the MLCS2k2 model \citep{Jha07} as described in Section \ref{sec:ABCwSDSS} with the following changes; we fix the cosmology to $\Omega_\Lambda = 0.73$, $\Omega_M = 0.27$ and $w=-1$, and we estimate photometric redshifts when we fit the light curves without using a host galaxy photo-z prior.\footnote{Please see Section 4.9 of the SNANA manual for details on measuring SN~Ia redshift from photometry \url{http://sdssdp62.fnal.gov/sdsssn/SNANA-PUBLIC/doc/snana\_manual.pdf}}  We use this sample to represent a realistic joint distribution between the spectroscopic and photometric redshifts.  We further assume that the spectroscopic redshift is equal to the true redshift $\ztru = \zspec$ and the observed redshift is the photometric redshift $\zobs = \zphot$.

\begin{figure}
\epsscale{0.8}
\plotone{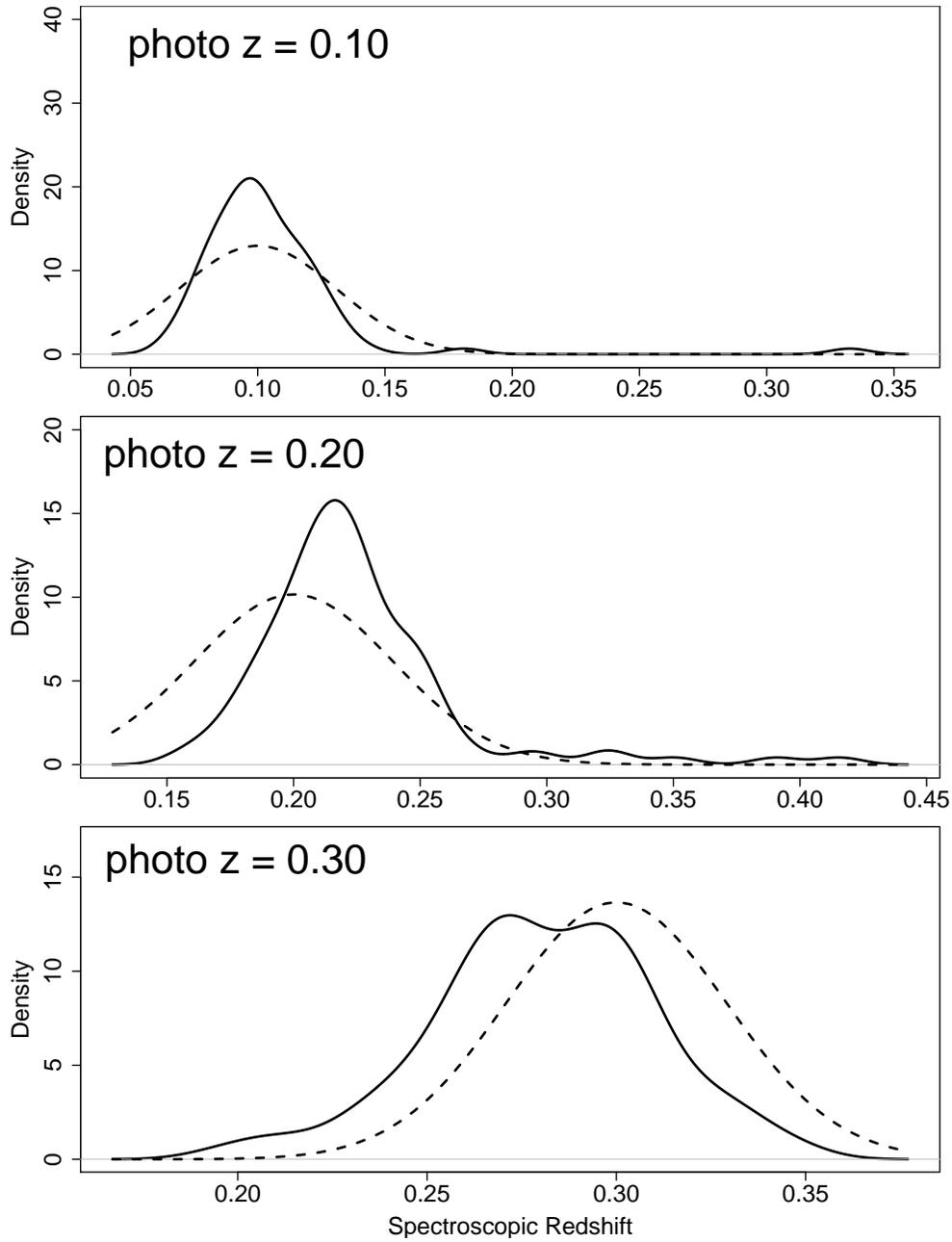}
\caption{Comparison between the assumed Gaussian joint distribution between $\zobs$
and $\ztru_i$ (dashed) and nonparametric fits (solid) through the simulated data shown in Fig.~\ref{fig:speczphotz}.  Three cross-sections are shown, one at each of photometric redshifts of
$0.1$, $0.2$, and $0.3$.  In each case, a bin of
width 0.02 is constructed, centered on these values, and the observations which fall into this bin are
used to estimate the distribution for spectroscopic redshift.
A Gaussian is not a terrible approximation to these cross-sections, but is it far from ideal.}
\label{fig:jointcross}
\end{figure}

\begin{figure}
\plotone{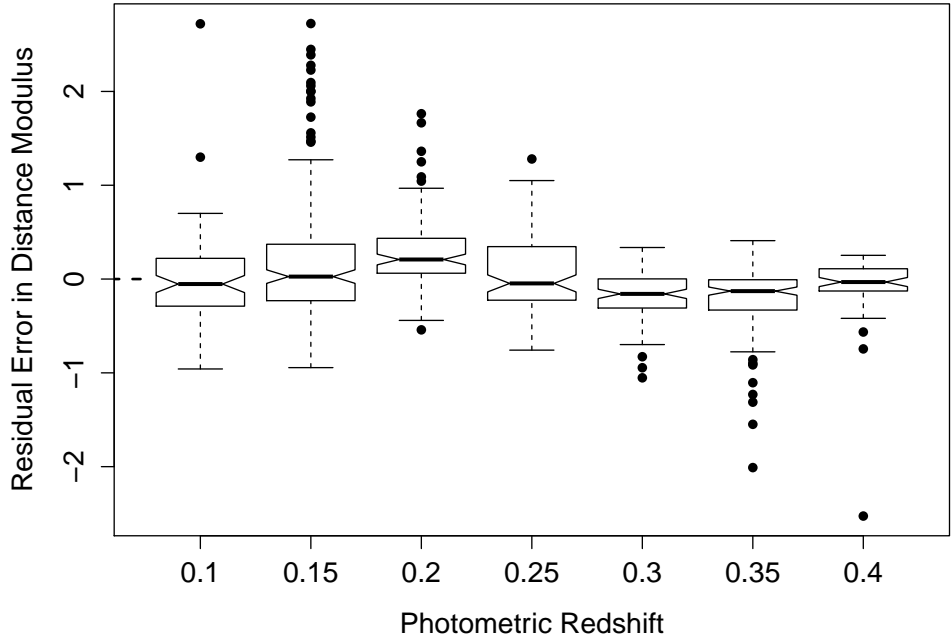}
\caption{Distance modulus residual defined as $\mu(\ztru_i,\theta)- \mu(\zobs_i,\theta)$ as a function of photometric redshift $\zobs_i$. Under the described Gaussian
approximation, these distributions should all have mean zero. The boxplots compare
the distribution in different narrow redshift bins. The top and bottom of the box indicate the 25th and 75th percentile, the center line marks the median, and the ``whiskers'' mark 1.5 times the inter-quartile range.  Points outside the whiskers are considered outliers. The ``notch'' in each boxplot
allows for comparison to determine statistical significance: if the notches of
two boxes do not overlap, then there is a statistically significant difference
between the medians of the populations. Hence, it is evident that there is
a bias introduced; the centers of these distributions are not always zero.
This bias indicates that the Gaussian 
model for the joint distribution of ($\ztru, \zobs$) 
is inappropriate.}
\label{fig:errorvsphotz}
\end{figure}

Fig.~\ref{fig:jointcross} shows three
cross-sections of the joint distribution of spectroscopic and photometric redshifts, 
comparing the photometric redshift distribution with the assumed Gaussian PDF. 
Our proposed model assumes that the horizontal cross-section of this
distribution at $\zphot$ is Gaussian with mean equal to $\zspec$.
This figure demonstrates that the Gaussian approximation 
to the distribution of $\zspec$ is not terrible.
Further, under this Gaussian approximation $\mu(\zphot_i,\theta)$ should
be approximately normal with mean $\tau_i$, i.e., under the linear approximation
the distance modulus estimated using the photometric redshift has mean equal 
to the true distance modulus.
Fig.~\ref{fig:errorvsphotz} uses boxplots to show 
the distribution of $\tau_i - \mu(\zphot_i,\theta)$ at various values
of $\zphot$ for the simulated data.
This plot reveals that there are significant deviations from the expected
difference of zero.

\begin{figure}
\plotone{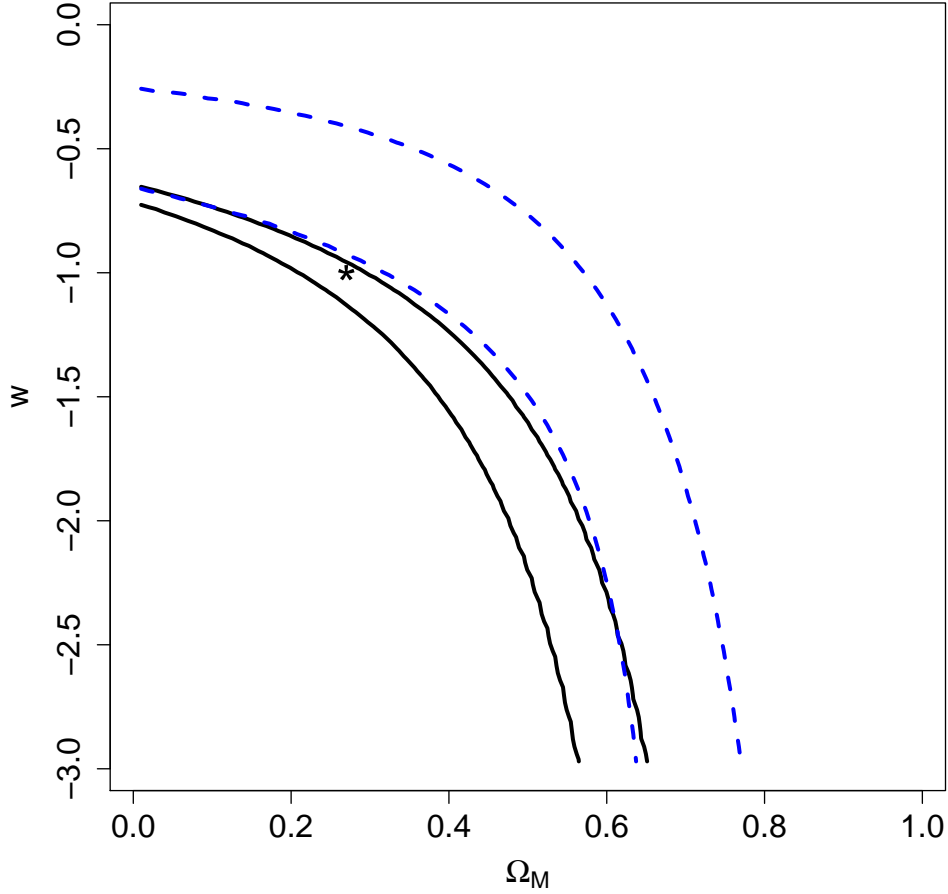}
\caption{Comparison between the 95\% credible regions for a simulated set of supernova formed by taking two approaches: (1) where the true redshift is known (black-solid line)
and (2) where the approximation described in Section 2.1 is utilized (blue-dashed line). 
The star is at the true value of the parameters used in the simulation. The increased width of the confidence region
is natural, given the use of photometric redshifts instead of spectroscopic redshifts, but the bias is a result of
the inadequacy of the assumed Gaussian model.}
\label{fig:compres2}
\end{figure}

The effect of this bias is made clear in Fig.~\ref{fig:compres2}. This figure
shows the 95\% credible region as constructed by two different methods, which
will be described below. In both cases, the data set utilized is the same.
To construct this data set
we simulated a sample of 200 SNe~Ia by drawing with replacement from the
$(\zspec, \zphot)$ sample shown in Fig.~\ref{fig:speczphotz}.  We then calculated 
$\tau = \mu(\zspec,\theta)$, where $\theta$ consists of $w=-1$ and $\Omega_M=0.27$ and assumed a flat universe.   
Finally the observed distance modulus $\mu$ is constructed by adding mean-zero Gaussian 
error onto $\tau$ with variance $\sigma_{\mu,i}^2 = 0.04$. 
The posterior
for $\theta$ is found for this dataset in two ways, and the 95\% credible region\footnote{The
region which comprises 95\% of the probability under the posterior is referred to as
a {\it credible region} to distinguish it from a frequentist {\it confidence region}.}
is displayed for each.
\begin{enumerate}
\item The solid line shows the credible region if the posterior is constructed using $\zspec$. It will serve as the fiducial reference for comparisons to the other region.
\item The dashed line is the credible region that results from using the
approximation described above, i.e., assuming that the observed distance modulus
has a Gaussian PDF with variance
\begin{equation}
\left[ \frac{\partial \mu(\zphot_i, \theta)}{\partial \zphot_i}\right]^2  \sigma^2_{\zphot,i} \, + \, \sigma^2_{\mu,i}.
\end{equation}
The point of emphasis here is that the additional uncertainty in the redshift is
now taken into account and reflected in the extra width of the region as compared
to the solid region. The shift from the solid region to the dashed region is the result of a bias.   
\end{enumerate}

The bias shown in Fig.~\ref{fig:compres2} is much like the {\it attenuation bias}
that results from inappropriately taking into account the errors in the predictor
variables in a regression setting: simply adding more error into the response
will not adequately account for this additional error. There are methods for
dealing with this additional error, but these are not practical in this setting
because of another fundamental challenge: the variance of the error in redshift
cannot be assumed to be constant, it needs to be modeled as a function of redshift.
This {\it heteroskedastic} error introduces significant obstacles to any method
that would seek to ``back out'' its effect on the estimates.
In the next section we will instead consider approaches that exploit our ability
to model and/or simulate the forward process that generated the data, and hence
allow us to incorporate in a natural way the errors due to the use of photometric redshifts.


\section{Approximate Bayesian Computation}
\label{sec:ABCmethod}

ABC methods simulate observations from the posterior distribution via algorithms that 
bypass direct calculation of the likelihood.  This is done by drawing model parameters from some distribution, generating simulated data based on these model parameters and reducing the simulated data to summary statistics.  Summary statistics are measures of the data designed to reduce the dimensionality of the data: they represent the maximum amount of information in the simplest form.  Model parameters that generate data sufficiently similar to the observed data are drawn from the posterior distribution.  This procedure allows one to simulate the complicated integral in Eq. \ref{intoverz} rather than evaluate it but instead relies upon the ability to simulate the forward process that generated the observed data.

Here we review two classes of ABC algorithms; ABC rejection samplers and adaptive ABC algorithms.  The roots of ABC techniques lie in the first class while the goal of adaptive ABC algorithms is to efficiently determine the relevant regions of parameter and probability space to sample from.  In this section we will adopt a Bayesian approach and endeavor to determine (approximately) the
posterior distribution of model parameters $\param$ given observed data $\dat$.
The posterior is given by 
\begin{equation} 
   \pi(\param \cond \dat) = \frac{\dens(\dat \cond \param)\pi(\param)}{\dens(\dat)},
\end{equation} 
where $\dens(\dat \cond \param)$ is the likelihood function and 
$\dens(\dat)$ is a normalization constant.  
For a review on ABC algorithms we refer the reader to \citet{Marin11}.

\subsection{ABC Rejection Samplers}

The basic ABC prescription is best considered for 
a situation in which the data $\dat$ are discrete:

\vspace{.1in}
\noindent
\fbox{
\begin{minipage}{0.93\linewidth}
\vspace{.1in}
\noindent {\bf Rejection Sampler: Discrete Case}
\begin{enumerate}
\setlength{\itemsep}{-6pt}
\item Draw candidate $\simparam$ from $\pi(\param)$
\item Simulate data $\simdat\sim\dens(\simdat \cond \simparam)$
\item Accept $\simparam$ if $\simdat = \dat$
\end{enumerate}
Repeat these steps until $N$ candidates are accepted.

\vspace{.05in}
\end{minipage}
}

\vspace{.1in}
\noindent Under this algorithm, 
the probability that $\simparam$ is accepted is exactly $\pi(\param \cond \dat)$.
Hence, it is simple in principle to generate a sample of size $N$
from the posterior distribution.  This sample is then used to estimate properties of the posterior distribution such as the 95\% credible region.

In practice, however, most data are continuous,
and we must instead decide to accept $\simparam$
if $\simdat$ is suitably ``close to'' $\dat$; 
hence, a metric or distance $\Delta(\simdat,\dat)$ must be chosen.
Under this setup, the accepted parameter vectors 
$\simparam$ are drawn from the posterior distribution conditioned on 
the simulated data being sufficiently close to the observed data. More precisely, 
the result will be a sample from the joint 
distribution $\dens(\dat,\param \cond \Delta(\dat,\simdat) \leq \epsilon)$ 
where $\epsilon > 0$ is a fixed tolerance.  
If $\epsilon$ is small and one marginalizes over $\dat$, 
then  $\dens(\param \cond \Delta(\dat,\simdat) \leq \epsilon)$ is 
a reasonable approximation to $\pi(\param \cond \dat)$ \citep{Sisson07}.  
Note that if $\epsilon$ is very large the sample will be effectively drawn from
the prior.
The continuous version of the ABC rejection sampler, 
introduced by \citet{Tavare97} and \citet{Pritchard99}, is built upon this idea:

\vspace{.1in}
\noindent
\fbox{
\begin{minipage}{0.93\linewidth}
\vspace{.1in}
\noindent {\bf Rejection Sampler: Continuous Case}
\begin{enumerate}
\setlength{\itemsep}{-6pt}
  \item Draw candidate $\simparam$ from $\pi(\param)$
  \item Simulate data $\simdat\sim\dens(\simdat \cond \simparam)$ 
  \item Accept $\simparam$ if $\Delta(\simdat,\dat) \leq \epsilon$
\end{enumerate}
Repeat these steps until $N$ candidates are accepted.

\vspace{.05in}
\end{minipage}
}

\vspace{.1in}
If the data have many dimensions, requiring that
$\Delta(\simdat,\dat) \leq \epsilon$ may be impractical.  For example, it would be nearly impossible to simulate $10^3$ supernovae to within $\epsilon$ of the observed data even with the correct cosmology due to random photometric error, let alone population variance in realizations of stretch and color distribution.
  
\citet{Fu97} and \citet{Weiss98} improved Step 3 by instead making the 
comparison between lower-dimensional summaries of the data; here these
will be denoted $\summ(\dat)$, or just $\summ$.
The ideal choice for $\summ$ would be a summary statistic that is 
a {\it sufficient statistic} for estimating $\param$. Technically, a
vector $\summ$ is sufficient if $\dens(\dat \cond \summ,\param)$ is not a function of
$\param$, and hence the posterior conditioned on $\summ$ is the same as the
posterior conditioned on $\dat$, i.e., $\pi(\param \cond \summ) = \pi(\param \cond \dat)$.
Of course, one cannot expect to derive an exactly 
sufficient statistic when the form of the likelihood
is not known. Hence, much current research in ABC is focused on the derivation of
{\it approximately sufficient statistics} or, more generally, summary statistics that
preserve important information regarding the parameters of interest.
\citet{Blum2012} provide an excellent overview and comparison of methods for
constructing summary statistics. These methods generally fall into two categories: 
those that sift through a list of candidate summary statistics to find the ``best'' summary statistic
as measured by some optimality criterion, and those that utilize the
ability to simulate data sets under different parameter values as part of a process of
fitting a regression where the responses are the parameters, and the predictors are
the simulated data.
This mapping is then used to 
transform observed summary statistics to parameters. 
For example, an early such example was \citet{Beaumont02}, 
who fit local linear regression to 
simulated parameter values on simulated summary statistics. 
The regression approach can be justified on theoretical grounds, see \citet{Fearnhead2012},
and \citet{Cameron12} used this approach for their astronomical application.
In our work, the relatively simple structure 
of the relationship
between the simulated data and the parameters of interest leads to a natural approach to
constructing a summary statistic: exploiting the known smooth distance modulus/redshift
relationship. In other applications, there will not exist such a simple one-dimensional
representation of the data, and these sophisticated approaches must be utilized.

There are advantages to the general ABC rejection sampler approach. Since 
each accepted parameter represents an {\it independent} draw 
from $\dens(\param \cond \Delta(\simsumm,\summ) \leq \epsilon)$, properties of 
the posterior distribution are easily estimated from the accepted sample. There
are no problems with such estimation due to dependence in the sample.
Also, the ABC rejection sampler is simple to code and trivial to parallelize.
However, the success of this method depends on how easy it is to simulate 
data from the model. If the model is complicated or if the acceptance 
rates are small, then the algorithm can be very expensive or inefficient. A low acceptance rate can be
caused by a diffuse prior relative to the posterior or by a poor choice for the summary statistic.
It is natural to consider approaches that do not rely upon independent sampling from the prior. In particular,
one would anticipate that it would be possible to ``learn'' from the parameter values that have been
accepted in the past to determine where good choices for future candidates $\simparam$.

\subsection{Adaptive ABC Algorithms}

The aforementioned challenges 
are the major motivations for the use of MCMC
techniques: instead of relying on random draws from a distribution to produce candidates, 
random walks are taken in parameter space.
\citet{Marjoram03} presented an MCMC version of ABC as follows:

\vspace{.1in}
\noindent
\fbox{
\begin{minipage}{0.93\linewidth}
\vspace{.1in}
\noindent {\bf ABC MCMC}

\vspace{.1in}
Initialize $\param_i,~i=1$\\
For $i$=1 to $i$=$N$ do: 

\vspace{.1in}
\begin{enumerate}
\setlength{\itemsep}{-6pt}
  \item Propose a move to $\simparam$ according to a transition kernel $q(\param_i \rightarrow \simparam)$
  \item Simulate $\simdat\sim\dens(\simdat \cond \simparam)$
  \item Measure $\simsumm$ from $\simdat$ 
  \item If $\Delta(\simsumm,\summ) \leq \epsilon$ proceed, else go to Step 1
  \item Set $\param_{i+1}=\simparam$ with probability
    \[
h(\param_i, \simparam)={\rm min}\left(1,\frac{\pi(\simparam)q(\param_i \rightarrow \simparam)}{\pi(\param_i)q(\simparam \rightarrow \param_i)}\right) 
\]
   and otherwise, $\param_{i+1}=\param_i$ 
  \item $i = i+1$
\end{enumerate}

\vspace{.05in}
\end{minipage}
}

\vspace{.1in}
\noindent Here $q(\param_i \rightarrow \simparam)$ is a proposal 
density, $h(\param_i, \simparam)$ is the Metropolis--Hastings 
acceptance probability and $N$ is the chain length.  The chain length 
is determined after meeting some convergence criterion (see, e.g., \citet{Cowles96}).
As is proved in \citet{Marjoram03}, the posterior distribution of interest 
$\pi(\param \cond \dat)$ is the stationary distribution of the chain.

The MCMC ABC algorithm can be much more efficient than the ABC rejection sampler, 
especially when the posterior and prior distributions are very different.
This efficiency is gained, however, at the cost of highly correlated $\param_i$. 
Additionally,
the MCMC ABC sampler can become inefficient if it wanders into a region of parameter 
space with low acceptance probability with a poor perturbation kernel.  
Successive perturbations have a small chance or being accepted and the chain 
can get ``stuck.''
It is worth noting that this algorithm is replacing the likelihood ratio
present in standard MCMC techniques with a one or zero based on whether or
not $\Delta(\simsumm,\summ) \leq \epsilon$. This is a significant
loss of resolution in the information that was present in the likelihood
ratio.


\citet{Sisson07} (improved upon by \citet{Beaumont09})
overcome the inefficiencies of a MCMC ABC algorithm via a method 
which they term {\it Population Monte Carlo} or {\it Sequential 
Monte Carlo} (SMC) ABC. 
The SMC ABC approach adapts the SMC methods developed 
in \citet{Moral06} to ABC. 
The algorithm learns about the target distribution 
using a set of weighted random variables that are propagated over iterations, similar to 
running parallel MCMC algorithms which interact at each iteration. 
The basic recipe of the SMC ABC algorithm is to initialize 
$N$ points in parameter space according to $\pi(\param)$.  
Points or particles are drawn from this sample, slightly perturbed, and are accepted for 
the next iteration if they meet the $\epsilon$ criterion. For each iteration, the tolerance 
$\epsilon$ is decreased, slowly migrating the $N$ particles into the correct region of parameter 
space when we have reached a pre-specified tolerance threshold.

\vspace{.1in}
\noindent
\fbox{
\begin{minipage}{0.93\linewidth}
\vspace{.1in}
\noindent {\bf SMC ABC}

\vspace{.1in}
Fix a decreasing sequence of tolerances $\epsilon = \epsilon_1, \epsilon_2,...,\epsilon_T$

\vspace{.1in}
For the first iteration, $t$=1:

\begin{quote}
\indent For $i$=1 to $i$=$N$ do:
\begin{enumerate}
\setlength{\itemsep}{-6pt}

\item Draw $\param_i^{t}$ from $\pi(\param)$
\item Simulate $\dat_i^{t}\sim\dens\left(\dat_i^t \cond \param_i^{t}\right)$
\item Measure $\summ^{t}_i$ from $\dat_i^{t}$
\item Proceed if $\Delta(\summ^{t}_i, \summ) < \epsilon_t$, else return to Step 1
\item Set $w_i = 1/N$
\item $i=i+1$
\end{enumerate}

Take $\tau_{t+1}^2$ equal to twice the weighted variance of the set $\{ \param_i^{t}:i=1,...,N \}$.

\end{quote}

\vspace{.1in}
For $t$=2 to $t$=$T$ do:

\begin{quote}

For $i$=1 to $i$=$N$ do:

\begin{enumerate}
\setlength{\itemsep}{-6pt}

\item Draw $\simparam$ from $\{ \param_j^{t-1}:j=1,...,N \}$ with probabilities $\{w_j^{t-1}\}$
\item Generate $\param_i^{t}$ from $K(\simparam, \tau_t^2)$
\item Simulate $\dat_i^{t}\sim\dens\left(\dat_i^t \cond \param_i^{t}\right)$
\item Measure $\summ_i^{t}$ from $\dat_i^{t}$
\item Proceed if $\Delta\!\left(\summ_i^{t}, \summ\right) < \epsilon_t$ else return to Step 1
\item Set   
\[
w_i^{t} \propto \frac{\pi(\param_i^{t})}{\sum_{j=1}^N w_j^{t-1} K\left(\tau_t^{-1}\!\left(\param_i^{t} - \param_j^{t-1}\right)\right)}
\]
\item $i=i+1$

\end{enumerate}

Take $\tau_{t+1}^2$ equal to twice the weighted variance of the set $\{\param_i^{t}:i=1,...,N\}$
\end{quote}
\vspace{.05in}
\end{minipage}
}

\vspace{.1in}
\noindent Here, $K(x)$ is a kernel which could be, e.g., a Gaussian kernel such that $K(x) \propto \exp{(-x^2/2)}$
and the weights are normalized after $N$ points have been selected. 
Following \citet{Beaumont09}, each particle is perturbed using a multivariate normal distribution
with mean centered on the particle's current position $\simparam$ and variance equal to twice 
the weighted empirical covariance matrix of the previous iteration $N(\simparam, \tau_t^2)$.  
Some work has been invested determining the most efficient method of perturbing points and includes 
implementing a locally adapted covariance matrix and incorporating an estimate of the 
Fisher information (see \citet{Filippi11}).

Since the target distribution is approximated by a random sample of $N$ particles that have migrated over iterations, 
properties of the posterior distribution are again properties of the sample, i.e., there is no covariance 
between the points as in the MCMC case.  Using the importance weighting scheme in \citet{Beaumont09} 
along with the distribution of particles in parameter space allows one to construct an estimate of
the posterior distribution and derive estimates of parameters of interest based on this posterior.  

SMC ABC has some distinct advantages over the other ABC methods.  Both the ABC rejection sampler and the MCMC ABC 
scheme become very inefficient when the tolerance is small.  SMC ABC derives its efficiency 
instead from sequentially learning about the target distribution by decomposing the problem into a 
series of simpler sub-problems.  The sequence of $\epsilon$'s can be chosen such that the acceptance 
rates are never too poor and the algorithm converges at a reasonable rate.  However, if the sequence 
of $\epsilon$ decreases too slowly the algorithm will be too computationally expensive and if it 
decreases too rapidly the acceptance rates will be too small.  An inefficient perturbation kernel will 
also result in a poor exploration of the space and similarly poor acceptance rates as many simulated datasets will be generated 
before $\Delta(\summ_i^t, \summ)<\epsilon_t$ is reached. 

ABC is an active field of research.  Recent improvements have been made by 
\citet{Barnes11}, who employ an information-theoretical framework to construct approximately sufficient statistics 
and \citet{Blum08}, who introduce a machine learning approach to estimate the posterior by 
fitting a nonlinear conditional heteroscedastic regression of the parameters on the summary statistics.  
The estimation is then adaptively improved using importance sampling.  For a review and study of the improvements 
made in ABC methods in recent years we refer the reader to \citet{Marin11}. 

\subsection{Example: Revisited}

Here, we apply SMC ABC to the stylized SN~Ia inference example introduced in \S~\ref{sec:SNeToy}.
The model is the same as was specified in that section. 
The ``observed data'' are simulated by constructing a sample of 200 SNe~Ia under a flat
cosmology with $\Omega_M = 0.27$ and $w = -1$. For this toy example, $H_0$ is assumed to be perfectly known as 72 km~s$^{-1}$~Mpc$^{-1}$.

\begin{figure}
\plotone{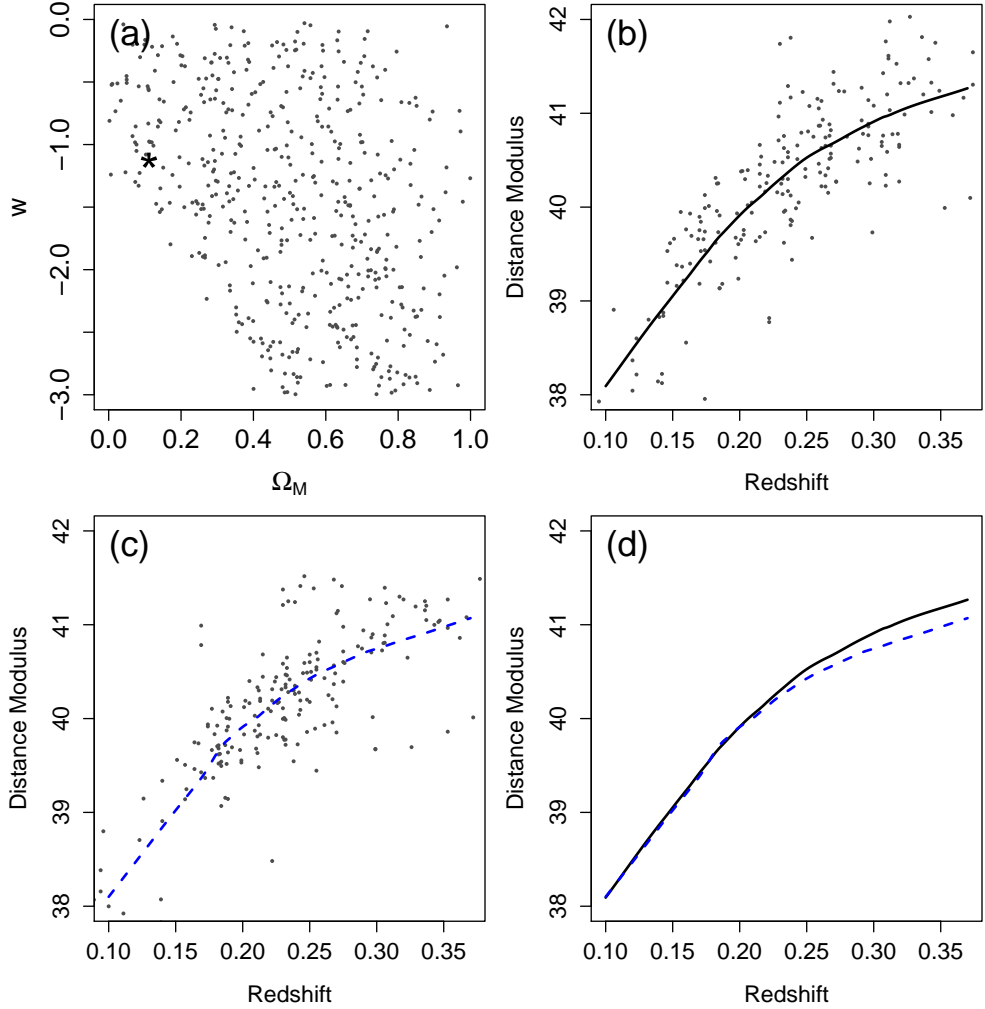}
\caption{Illustration of key steps of the SMC ABC algorithm in the example. Panel (a): a collection of 500 particles plotted in the relevant parameter space from an intermediate iteration of the SMC ABC algorithm.  A random particle is selected, plotted as the star, and perturbed a small amount.  Panel (b): the simulated dataset corresponding to the perturbed particle from panel a.  The line is a nonparametric smooth of the data and represents the summary statistic. Panel (c): ``Observed'' data.  The dashed line represents a nonparametric smooth of the observed data.  Panel (d): a comparison between the simulated and observed datasets via 
the sum of squared deviations across the length of the curve. The particle is accepted in this iteration even though the curves are discrepant at high redshift as the tolerance is not small enough to reject it.  Such a point would likely be rejected in a future iteration as the tolerance is decreased (see Fig.~\ref{fig:toyphot}).}
\label{fig:tutorial}
\end{figure}

Fig.~\ref{fig:tutorial} depicts key steps in the SMC algorithm as applied to this situation.  The prior is chosen to be uniform over the region $0 < \Omega_M < 1$ and $-3 < w < 0$.
A collection of 500 $(\Omega_M, w)$ pairs, often called {\it particles} in the context of SMC methods, is migrated through the iterations of the algorithm.
Fig.~\ref{fig:tutorial}a shows the collection of 500 particles at the conclusion of one of the early time steps.
One of these particles is chosen at random and perturbed a small amount;
the parameter combination is $\Omega_M = 0.11$ and $w = -1.21$, and is shown as the star in the
plot. This parameter combination in denoted $\param_i^{t}$ in the algorithm above.
Simulated data are created by drawing a collection of 200 $(\zobs, z')$ pairs, sampling with replacement, from the collection shown in Fig.~\ref{fig:speczphotz}.
With $\param_i^t$ specified and the 200 true redshifts, it is trivial to calculate the distance
modulus of each SN~Ia, and then add uncertainty using a Gaussian PDF with variance $ \left ( \sigma_{\mu,i} \right )^2 = 0.04$.
Fig.~\ref{fig:tutorial}b shows the resulting simulated distance moduli plotted against the photometric
redshifts $\zobs$. The point is that this is a plot that can be created using observable
data: these data comprise the $\dat_i^{t}$ that appear in the algorithm above.

A key step in any implementation of SMC ABC is the choice of the summary statistic.
Here, the summary statistic $\summ_i^{t}$ is found by applying a nonparametric regression
smoother through these data; this curve is shown in Fig.~\ref{fig:tutorial}b. (The approach
used to perform this smooth is briefly presented in the Appendix.)
The motivation for this choice is as follows: as stated above, ideally we would choose
a sufficient statistic as our summary statistic. A sufficient statistic is a summary that 
separates out from the full data that portion which is useful for estimating $\param$.
In this case, we know that the relationship between redshift and distance modulus for fixed
$\param$ is a smooth curve. The deviation of the data from a smooth curve can be solely
attributed to random error in the measurements, error which is not all informative of the
value of $\param$. For this reason, it is reasonable to believe that a smoothed version of
the points shown in Fig.~\ref{fig:tutorial}b captures all of the useful information for estimating $\param$.

\begin{figure}
\plotone{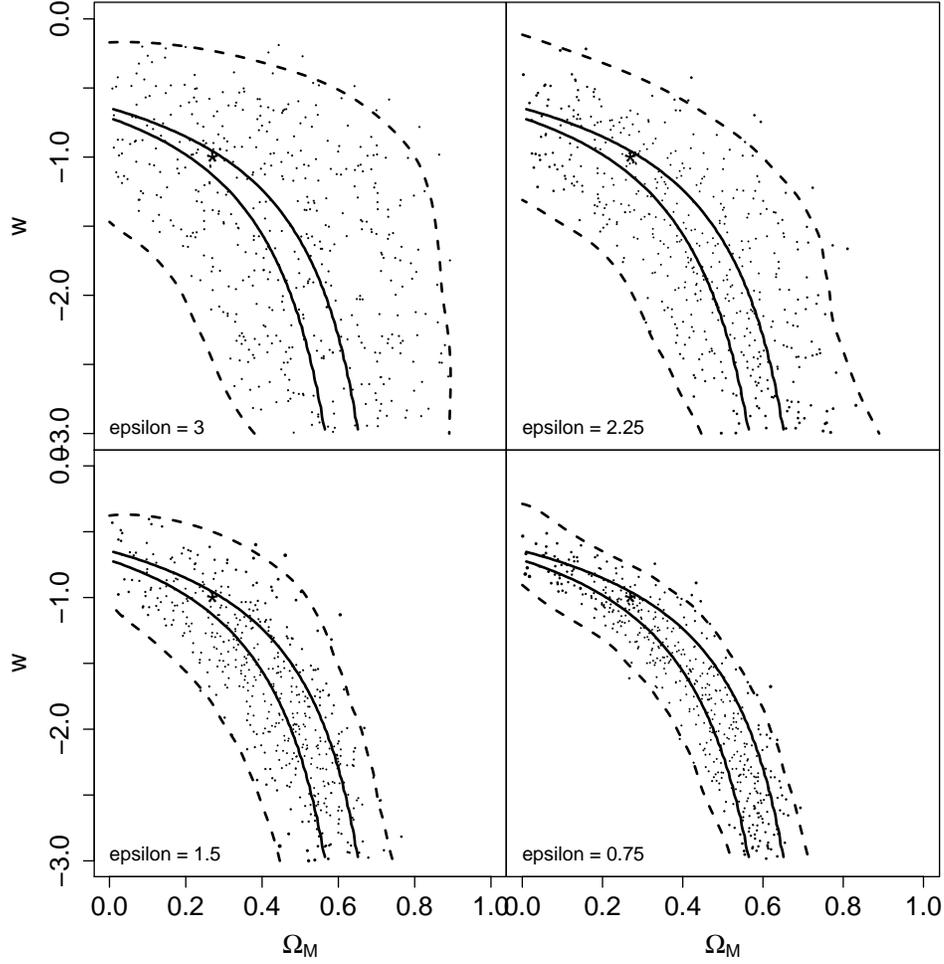}
\caption{Progress of the ABC SMC algorithm in estimating the posterior distribution for the toy example. As $\epsilon$ decreases, the
collection of particles converges to a sample from the posterior (when the weights are taken into account). The solid contour
is the 95\% credible region that would have been formed by someone who had knowledge of the spectroscopic redshifts. 
The dashed contours result from fitting to the output of the ABC algorithm.
Compare
with Fig.~\ref{fig:compres2} to note the reduction of the bias that resulted from the Gaussian approximation.
Note that it is not expected that these contours will be the same, as the ABC simulations are built upon
data using photometric redshifts; hence, there is additional uncertainty in the parameter estimates.}
\label{fig:toyphot}
\end{figure}

The comparison between the real and simulated data will be done via these smooth curves.
Fig.~\ref{fig:tutorial}c shows the observed data, along with the result $\summ$
of applying the same smoothing procedure to these data.
Finally, in Fig.~\ref{fig:tutorial}d, these two curves are compared via a simple 
distance calculation
between these curves, namely, the sum of squared deviations across the length of the
curve. The particle is accepted in this iteration, because even though the
curves differ at high redshift, the tolerance is not sufficiently small yet to reject at
this difference.
Fig.~\ref{fig:toyphot} shows how the collection 500 particles evolves over the
steps of the algorithm. As the steps progress, the particles converge in and approximate
a sample from the posterior. The notable feature of this result is that this posterior
is centered on the solid contours. Just as in Fig.~\ref{fig:compres2}, these contours represent
the posterior as derived by someone who had full knowledge of the redshifts. It is clear
that by avoiding the unjustified Gaussian assumptions made in \S~\ref{sec:SNeToy}, the
bias that was present in the previous posterior based on photometric redshifts has been removed.




\section{SMC ABC Cosmology with SDSS-II Supernovae}
\label{sec:ABCwSDSS}


In this section we apply SMC ABC to first year data from the SDSS-II Supernova Survey~\citep{Holtzman08,Kessler09b}.  The development of the sophisticated supernova simulation and analysis software SNANA~\citep{Kessler09a} has made possible the comparison between the SDSS-II supernova sample and simulated data sets and is a natural first choice to test ABC methods in cosmology.  
The purpose of this section is to demonstrate that ABC can be used to estimate an accurate posterior distribution.  We use the spectroscopically confirmed sample to estimate cosmological parameters from assuming a spatially flat universe and a constant dark energy equation of state parameter, $w$.  In this section we discuss how we create simulated data sets, our ABC setup, and compare our posterior distributions for the matter density $\Omega_M$ and the equation of state parameter $w$ with those from a $\chi^2$ analysis using statistical errors only.  We close this section demonstrating the full utility of ABC by including Type IIP supernovae contamination to the SDSS sample and estimating the correct posterior distribution with ABC.

\subsection{Simulation Setup}\label{sec:setup}


For this analysis we will use data from the fall 2005 SDSS-II Supernova Survey which were published in \citet{Holtzman08}.  For detailed information regarding the scientific goals and data processing for the survey we refer the reader to \citet{Frieman08}, to \citet{Sako08} for details of the supernova search algorithms and spectroscopic observations and to Section 2 of \citet{Kessler09b} for a brief summary of the survey.

Our goal is to compare the derived posterior distributions for $\Omega_M$ and $w$ using ABC with those from \citet{Kessler09b} which were done using a more traditional $\chi^2$ analysis.  To make this comparison as meaningful as possible we apply the same relevant selection cuts to the data.  Therefore, defining $t_0$ as the time of peak brightness in rest-frame B according to MLCS2k2 such that $t-t_0=0$, we require for each SN~Ia light curve, one measurement before peak brightness and one measurement more than 10 days after peak brightness.  Additionally we require five measurements with $-15 < t-t_0 < 60$~days.  These requirements ensure adequate time sampling to yield a robust light-curve model fit.  \citet{Kessler09b} additionally require one measurement in $gri$ with a signal-to-noise greater than 5 to put a floor on the quality of data and require $P_{\rm fit} > 0.001$, where $P_{\rm fit}$ is MLCS2k2 light-curve fit probability based on $\chi^2$. This requirement is designed to remove obvious peculiar SNe~Ia in an objective fashion.  

All supernovae in this sample have unambiguous spectroscopic confirmation and we use photometry in $g$, $r$, and $i$ bands.  This leaves us with 103 SDSS SNe~Ia.  This sample is identical to \citet{Kessler09b}'s sample A and can be taken from their Table 10.

We can broadly separate the treatment of variables in the likelihood into two categories: (1) those which are of cosmological interest and (2) nuisance parameters.  One will be able to construct posterior distributions for all parameters in the first category, in this case $\param  = [ \Omega_M, w ]$, while sampling from the probability space spanned by the set of nuisance parameters when generating simulated data sets.   

We use SNANA to simulate sets of supernovae from different cosmologies.  The idea is to randomly sample from the probability distributions of each nuisance parameter every time a simulated set of supernovae is generated.  If we were to fix the cosmology and simulate many data sets, the probability space spanned by the nuisance parameters should be reflected in the variance of the sets of simulated data.

Within SNANA we will use the MLCS2k2 model \citep{Jha07} to simulate SN~Ia light curves.  We use the same modified version of MLCS2k2 that was developed and trained in \citet{Kessler09b}.  In this model the observed model magnitudes corrected for Galactic extinction, $K$-correction, and time dilation, for each passband, $X$, are given by
\begin{eqnarray} 
{\bf m}_X(t-t_0) &=& {\bf M}^0_X + \mu_0 + {\bf \xi}_X \left ( \alpha_X + \frac{\beta_X}{R_V} \right ) A_V^0  \nonumber \\
&& + {\bf P}_X\Delta  + {\bf Q}_X \Delta^2 \label{eq:mlcs2k2}
\end{eqnarray}
where ${\bf M}_X^0$ are the fiducial absolute magnitudes, $\mu_0$ is the distance modulus, $R_V$ and $A_V^0$ are the host galaxy extinction parameters, and ${\bf P}_X$ and ${\bf Q}_X$ describe the change in light-curve shape as a quadratic function of $\Delta$.  Quantities that are functions of phase are in bold. ${\bf M}_X^0$, ${\bf P}_X$, and ${\bf Q}_X$ are estimated from a training set leaving $t_0$, $\mu_0$, $\Delta$, $A_V^0$, and $R_V$ as the free parameters. 

The distance modulus can be related to the luminosity distance for a flat universe with a constant dark energy equation of state parameter of $w=-1$ in the following way
\begin{eqnarray}
\mu_0 &=& 5 \log\left ( d_L / 10 {\rm pc}\right ) \\
&=& 5 \log\left ( c(1+z) \int_0^z \left [ \Omega_M(1+z^{\prime})^3 + \Omega_{\Lambda} \right ]^{1/2} dz^{\prime} \right ) \nonumber \\
&& - 5 \log{H_0 + 25.} \label{eq:lumdist}
\end{eqnarray}
Note that a change in $H_0$ simply scales the distance modulus. It is easy to see that if one rewrites Eq. \ref{eq:mlcs2k2} in terms of luminosity distance that a degeneracy arises between $H_0$ and $M_V$.  Even if $H_0$ is known from some other experiment, $M_V$ would still need to be marginalized over.

$\xi_X$ is defined as
\begin{equation} \xi_X = \frac{A_X}{A_X^0}\end{equation}
and is equal to unity at maximum light.  
This framework allows one to separate out the time dependence of the extinction while being insensitive to the total extinction $E(B-V)$ and the extinction law $R_V$.  

A major advantage of MLCS2k2 is that it allows one to separate reddening resulting from dust in the host galaxy (third term in Eq. \ref{eq:mlcs2k2}) from intrinsic color variations of the supernova which are captured by $\Delta$.  The validity of this approach depends on how separable these two terms are, how well intrinsic color is predicted by light curve shape, and relies on accurate models of the distribution of extinction with redshift \citep{Wood-Vasey07}. 

To generate a simulated set of data, we assume a flat universe and choose $\Omega_M$ and $w$ from flat priors over the range $[0,1]$ and $[-3,0]$ respectively.  One could instead draw cosmological parameters from priors based on the SDSS detection of the baryon acoustic oscillations \citep{Eisenstein05} and the five-year Wilkinson Microwave Anisotropy Probe observations (WMAP-5) of the cosmic microwave background \citep{Komatsu09}.  A random supernova redshift is selected from a power law distribution given by $\frac {dn}{dz} \sim (1+z)^{\beta}$ where $\beta = 1.5 \pm 0.6$ \citep{Dilday08}.  $\Delta$ and $A_V$ are then drawn from empirical distributions determined in Section 7.3 of \citet{Kessler09b}.  Using the parameterization of \citet{Cardelli89} to describe the extinction with $R_V=2.18$ (as determined from Section 7.2 in \citealt{Kessler09b}), the MLCS2k2 light-curve model can now be used to generate supernovae magnitudes which are then $K$-corrected using spectral templates from \citet{Hsiao07} into observer frame magnitudes.

SNANA then chooses a random sky coordinate consistent with the observed survey area and applies Galactic extinction using the \citet{Schlegel98} dust maps, chooses a random date for peak brightness, and selects observed epochs from actual SDSS survey observations. Noise is simulated for each epoch and filter and includes Poisson fluctuations from the SN~Ia flux, sky background, CCD read noise, and host galaxy background. 

The simulation allows one to add additional intrinsic variations in SN~Ia properties to better match the observed scatter in the Hubble diagram.  We do this by ``color smearing.''  A magnitude fluctuation drawn from a Gaussian distribution is added to the rest-frame magnitude for each passband leading to a change in model colors of $\sim 0.1$~mag.  SNANA also includes options to model the search efficiency of the survey. 

The aforementioned selection cuts on the observed data are then applied to the simulated data.  This process is done for a selected cosmology for $\sim 100$ SN~Ia  over the redshift range of $[0.02, 0.45]$, similar to the SDSS data, assuming a redshift uncertainty of 0.0005.  Finally, the distance modulus is measured by performing an MLCS2k2 light curve fit assuming the same prior on $A_V$ and $\Delta$ from which the data were simulated. 

\begin{figure}
\plotone{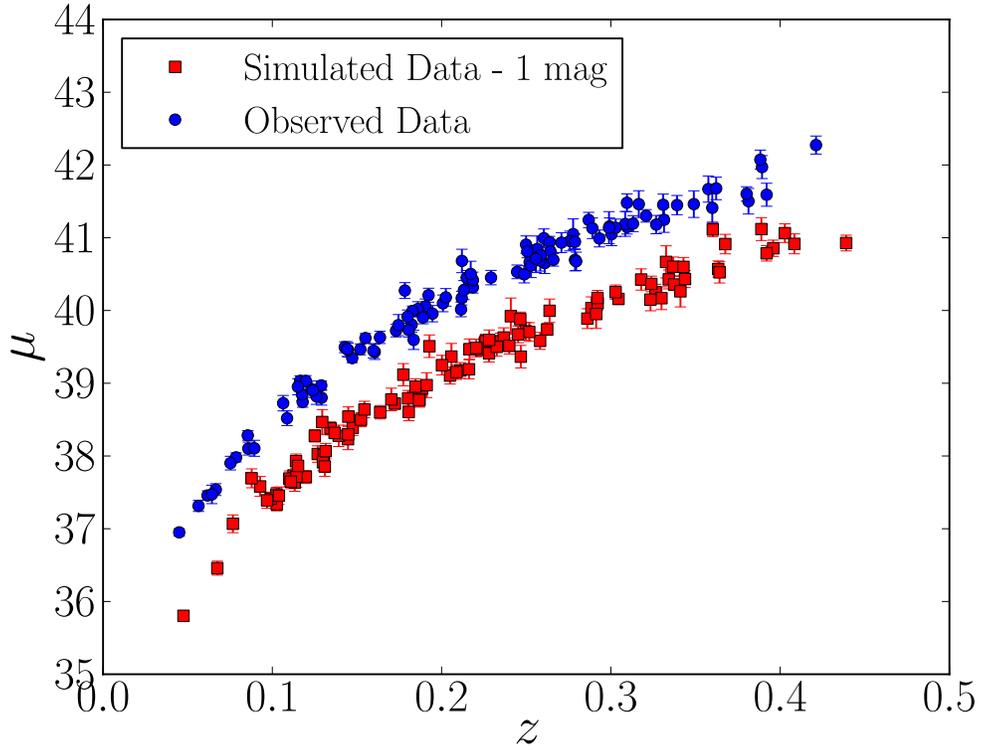}
\caption{Hubble diagram for the observed data in blue and a simulated dataset in red.  The simulated dataset is offset from the observed data by 1~mag and was generated assuming $\Omega_M=0.3$ and $w=-1.0$.  The distance modulus uncertainties, intrinsic scatter, and redshift distributions are well reproduced in the simulated dataset.  Simulated datasets like this one with different cosmologies are used in our ABC analysis.}
\label{fig:comparedatasets}
\end{figure}

In Fig.~\ref{fig:comparedatasets} we plot the distance modulus as a function of redshift for the SDSS data in blue and a simulated data set in red.  For the simulated data set we assume that $\Omega_M = 0.3$ and $w=-1.0$.  
The simulated data have been offset by 1~mag for clarity.  The distance modulus uncertainties, intrinsic scatter, and redshift distributions are similar between the simulated and observed data sets.  

\subsection{SMC ABC Implementation}

\begin{figure}
\plotone{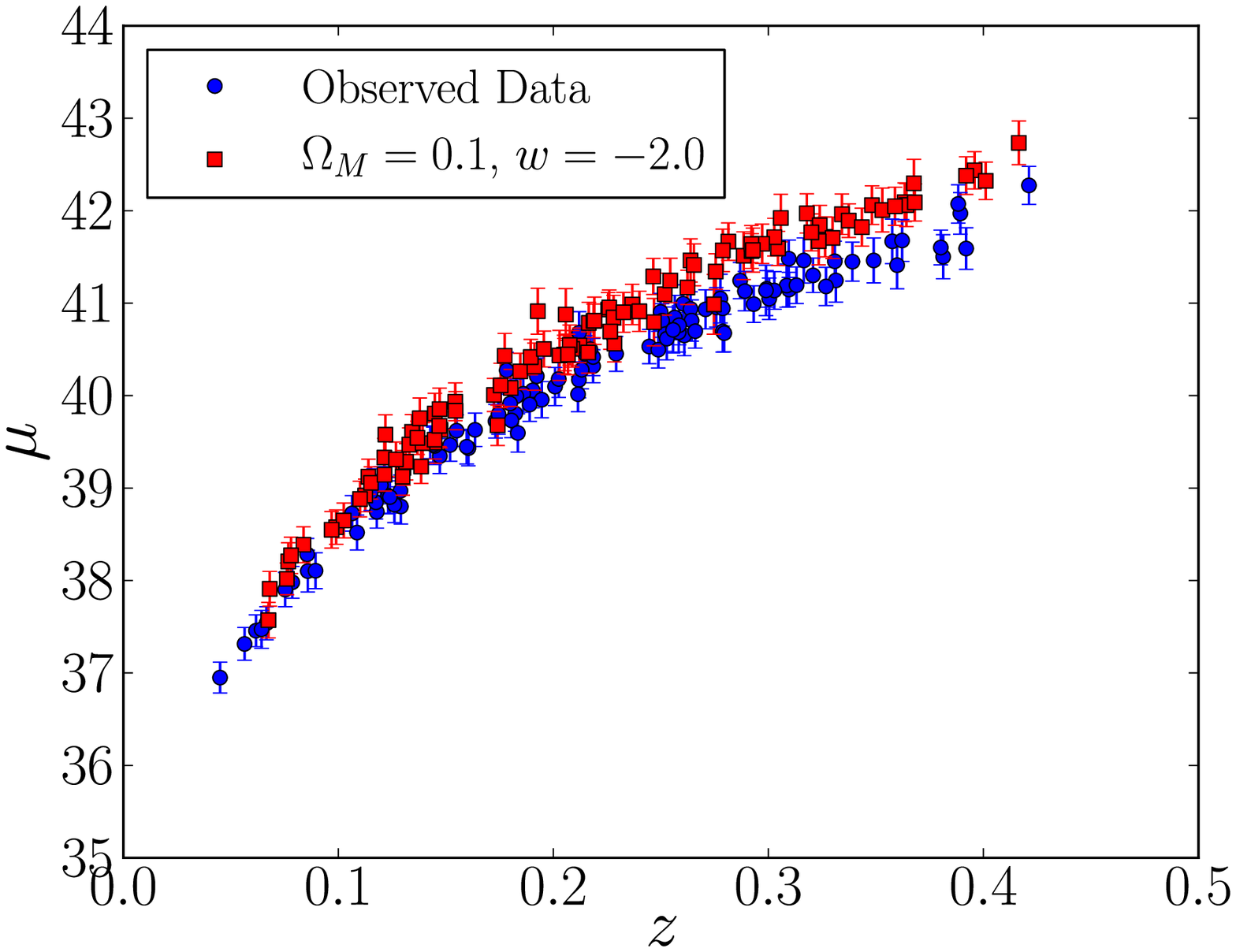}
\plotone{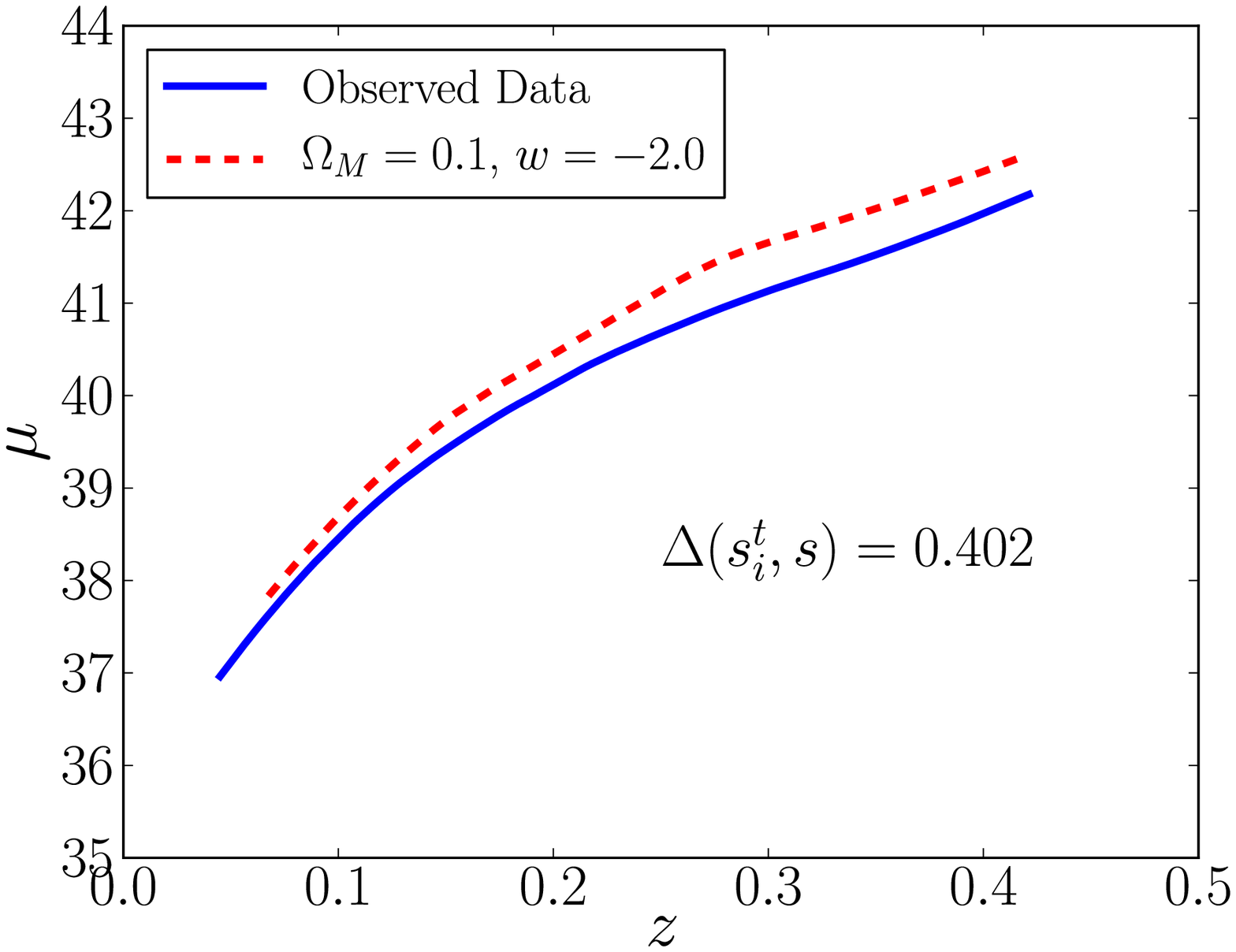}
\caption{Illustration of the distance metric using SDSS data and simulated data from SNANA.  Top: Hubble diagram for the observed data in blue and a simulated dataset in red.  The simulated data were generated assuming $\Omega_M=0.1$ and $w=-2.0$.  Bottom: Nonparametric smooth of the two datasets.  The distance metric is defined to be the median absolute deviation between the smoothed curves which is equal to 0.402 for this case.  Our final tolerance is 0.033. }
\label{fig:DeltaExample}
\end{figure}

To calculate the measure of similarity between the observed and simulated data sets, $\Delta(\summ^{t}_i, \summ)$, we turn to the Hubble diagram.  In the top panel of Fig.~\ref{fig:DeltaExample} we show $\mu$ versus $z$ for our observed data and a simulated data set with $\Omega_M=0.1$, and $w=-2.0$.  
A reasonable distance measure could be the Euclidean distance between the data sets at the redshifts of the observed data. However, in keeping with the notion of summary statistics, we would like to compare a smooth representation of the two data sets rather than the data themselves.  In the bottom panel of Fig.~\ref{fig:DeltaExample} we show a non-parametric smooth of the simulated and observed data.  The details on how we perform the non-parametric smooth are in Appendix \ref{appendix:smooth}.  We opt for a non-parametric smooth in the interests of efficiency and to prevent inserting additional assumptions about the data in an intermediate step
in contrast to fitting the data with a cosmology fitter.  We now define $\Delta(\summ^{t}_i,\summ)$ to be the median absolute deviation between the smoothed data sets evaluated at the observed redshifts.  We choose this because it is simple, it is robust to poor smoothing at high and low redshifts, and allows for a physical interpretation of the minimum tolerance.  Since we are basically measuring the distance between the two data sets in distance modulus, we consider our minimum tolerance to be equal to the median uncertainty in the smoothed observed data, i.e., we declare the observed data and simulated data sets sufficiently similar when the simulated data are within the error of the observed data. 

For simplicity in this analysis we fix the value of $H_0$ to 65~km~s$^{-1}$~Mpc$^{-1}$ to restrict the relevant region of parameter space.  This improves the efficiency of the ABC algorithm and more importantly, makes the comparison between ABC and $\chi^2$ more striking.  However, we note that for this particular definition of the distance metric the simulated value of $H_0$ directly scales $\Delta(\summ^{t}_i, \summ)$ in a trivial manner.  One could naively treat $H_0$ as a nuisance parameter and randomly sample $H_0$ from a flat prior over some range.  Since $H_0$, $w$, and $\Omega_M$ are correlated, a faster approach would be to add $H_0$ as another cosmological parameter, adding a third dimension to the parameter space.  The particles would then trace out the three-dimensional posterior distribution from which one could marginalize over $H_0$ to obtain the two-dimensional projection.  Given the simulation expense, one would like to take advantage of the simple relationship between $H_0$ and $\Delta(\summ^{t}_i, \summ)$.  To this end one could calculate a set of $\Delta(\summ^{t}_i,\summ)$s corresponding to a range of Hubble parameter values for a given $w$ and $\Omega_M$.  The particle is then accepted with a percentage based on the number of $\Delta(\summ^{t}_i, \summ)$ elements that meet the tolerance criterion.  This avoids re-simulating data sets a given number of times over a range of $H_0$ values while still sampling the probability space fully and thus marginalizing over $H_0$.   

We choose $\epsilon_t$ according to the distribution of $\{ \Delta(\summ^{t-1}_i, \summ):i=1,...,N \}$ instead of having a predefined sequence of tolerances to walk though.   
For the first iteration, we accept all points, i.e., the tolerance is infinite.  For the next iteration, $t=2$, the tolerance $\epsilon_{t=2}$ is set to the 25\% percentile of $\{\Delta(\summ^{1}_i, \summ)\}$.  All subsequent $\epsilon$s are the 50\% percentile of the previous iteration.   A percentile which is too large allows for many acceptances and will not localize into the correct region until $T$ is large.  Conversely, if one is too strict in their sequence of tolerances,  many simulations are required before a point is accepted.  We found that putting a stricter cut on what $\epsilon$ should be early on helps concentrate quickly into the correct area of parameter space, requiring fewer simulations in future iterations. 

We define $\epsilon$ to be sufficiently small when it is less than the uncertainty on the non-parametric smooth of the observed data, which we estimate via bootstrap.  The median uncertainty on the non-parametric smooth for the SDSS data set is 0.033. We require $\Delta(\summ^{t}_i, \summ)$ for each particle to be less than this value at the final iteration.

We choose $N=150$ particles and run the code on eight different processors.  As the initial particles are independently drawn between the three runs, the results can be combined to better estimate the posterior distribution.  However, the sequence in $\epsilon$ is slightly different for each run.  In practice one should parallelize the code at the level of accepting $N$ points so that there is just one sequence of tolerances.  Ours do not vary significantly and is not a concern for our demonstration.  

Properties of the posterior distribution are then drawn from the final sample of particles and their weights which meet the minimum tolerance criteria.

\subsection{Results and Discussion}

It is useful to first review the cosmological analysis performed in \citet{Kessler09b}.  MLCS2k2 provides an estimate of the distance modulus for each supernova.  The $\chi^2$ statistic is then calculated over a grid of model parameters and used to derive cosmological parameter estimates.  Recall that $-2 \ln(\pi(\param \cond \dat)) = \chi^2$.  The $\chi^2$ statistic for the SDSS supernova sample is calculated according to 
\begin{equation} \chi^2 = \sum_i \frac{ \left [ \mu_i - \mu_{\rm mod}(z_i \cond w, \Omega_M, H_0) \right ]^2 } {\sigma_{\mu}^2} \end{equation}
where $\mu_i$ and $z_i$ are the distance modulus returned from MLCS2k2 and measured redshift of the supernova, and $\mu_{\rm mod}$ is the model magnitude.  
The distance modulus uncertainties are given by
\begin{equation} \sigma_{\mu}^2 = \left ( \sigma_{\mu}^{\rm fit} \right )^2 + \left ( \sigma_{\mu}^{\rm int} \right )^2 + \left ( \sigma_{\mu}^z \right )^2 \label{eq:sigmamu} \end{equation}
where $\sigma_{\mu}^{\rm fit}$ is the statistical uncertainty reported by MLCS2k2, $\sigma_{\mu}^{\rm int}=0.16$ is additional intrinsic error, and 
\begin{equation} \sigma_{\mu}^z= \sigma_z \left ( \frac{5}{\ln{10}} \right ) \frac{1+z}{z(1+z/2)}. \end{equation}
The posterior distributions for $\Omega_M$ and $w$ assuming a flat universe can then be found by marginalizing over $H_0$.  Recall for our comparison that we are fixing the value of $H_0$ and do not need to marginalize over $H_0$.

\begin{figure}
\epsscale{0.50}
\plotone{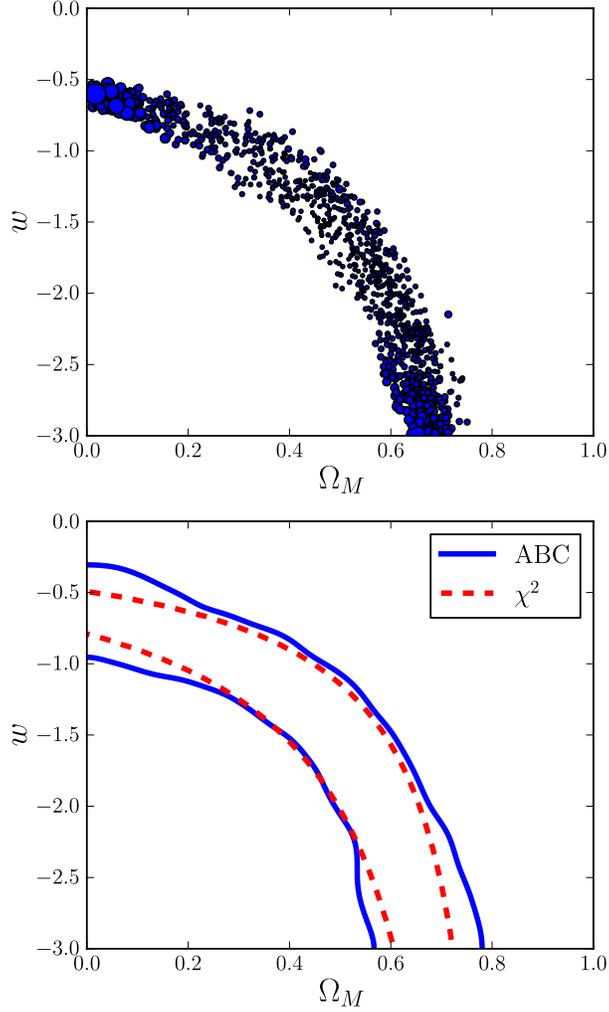}
\caption{Comparison of SMC ABC with a $\chi^2$ analysis.  Top: Particles from the final ABC iteration.  Bottom: The 95\% credible regions from ABC (blue-solid) and $\chi^2$ (red-dashed).  The contours between ABC and $\chi^2$ are well matched except near the boundaries.  The discrepancy results from the sharp boundaries of our prior.  ABC is attempting to account for the fact that there is relevant parameter space which it cannot explore.}
\label{fig:results}
\end{figure}

In Fig.~\ref{fig:results} we compare our posterior distribution to that found using the approach described above.   The top plot shows the particles from the final iteration of the SMC ABC algorithm.  The area of the particle symbol represents the weight.  These points and their weights represent a sample from the posterior distribution.  
We estimate the 95\% credible region from this sample and compare with the 95\% confidence region from a $\chi^2$ analysis in the bottom plot.  
Overall the contours are well matched.  The weights on the particles become large just inside the hard boundaries set by the priors on $\Omega_M$ and $w$. 
The algorithm is accounting for the fact that there is parameter space beyond the boundary which it cannot explore.  
This is similar to an MCMC algorithm running into a boundary and sampling more in that region because it cannot cross the boundary.  As a result the ABC contours become wider than those from $\chi^2$ near the boundaries.

We reiterate that the goal of this exercise was not to derive new cosmological constraints but merely to see how well we can recover the likelihood contours presented in \citet{Kessler09b} using a simple implementation of SMC ABC.  We demonstrate that we can recover the posterior distribution derived from current analysis techniques with the hope of convincing the reader this approach will be useful in the near future.  
We do note that the A in ABC stands for ``Approximate.''  One should expect slight differences in the estimated posterior distributions due to choices of distance metric, summary statistics, and final tolerance.

\subsection{Type IIP Contamination}

\begin{figure}
\plotone{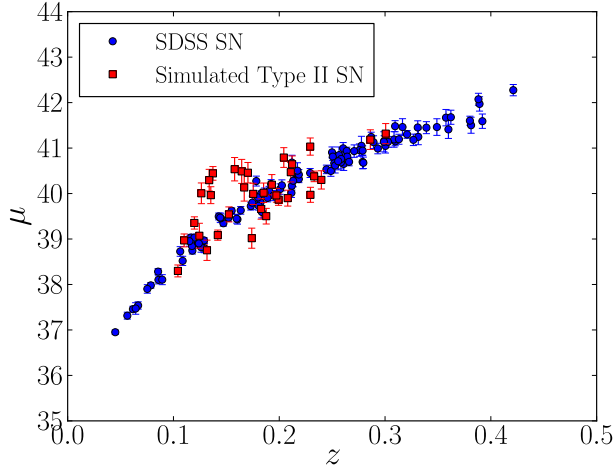}
\caption{SDSS sample plus 34 Type IIP supernovae simulated with SNANA.  This combined dataset is our ``observed'' sample for the type contamination analysis.}
\label{fig:typeconthd}
\end{figure}

We add 34 simulated Type IIP supernovae to the SDSS sample so that the overall type contamination is 25\%.  While the amount and type are a bit extreme it is useful for illustrative purposes.  We use SNANA to simulate the data which uses spectral templates and smoothed light curves of well observed supernovae.  We use the ``NONIa'' option which computes the observer magnitudes from the spectral energy distribution and we set MAGOFF=-0.6 and MAGSMEAR=0.9.  For details on these keywords and additional information on simulating non-Ia light curves we refer the reader to Section 3.5 of the SNANA manual.\footnote{\url{http://sdssdp62.fnal.gov/sdsssn/SNANA-PUBLIC/doc/snana\_manual.pdf}}  
The selection cuts, other observing parameters, and fitting procedure remain as described in Section \ref{sec:setup}.  Our new sample is plotted in Figure \ref{fig:typeconthd}.

We modify our SMC ABC analysis as follows; after drawing cosmology parameters from $\pi(\param)$, we simulate and fit additional Type IIP light curves in the aforementioned manner and add those to our simulated Type I data.  From this point the SMC ABC algorithm proceeds as before.  Our new final tolerance has increased to 0.038 due to the additional scatter in the Hubble diagram.  

\begin{figure}
\plotone{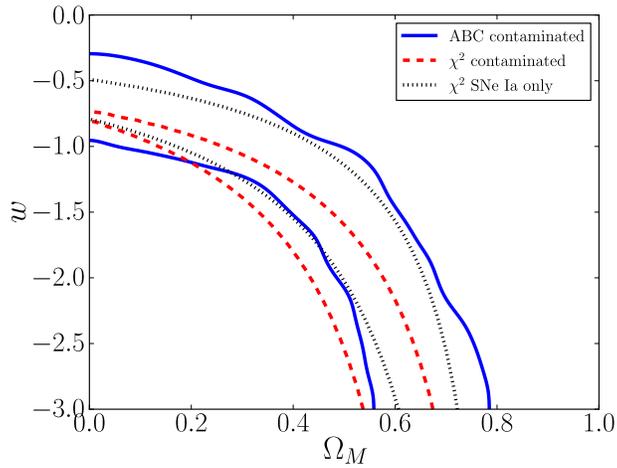}
\caption{95\% credible region from ABC (blue-solid) and the 95\% confidence interval from $\chi^2$ for the SDSS sample with type contamination (red-dashed) and the original SDSS sample (black-dotted).  The type contamination biases the $\chi^2$ result.  ABC reproduces the entire credible region without this bias and reflects additional uncertainty due to increased scatter in the Hubble diagram.}
\label{fig:resultstypeII}
\end{figure}

The resulting 95\% credible region is plotted in Fig. \ref{fig:resultstypeII} as the blue-solid line along with the 95\% confidence regions from $\chi^2$ with (red-dashed) and without (black-dotted) type contamination.  The contours from the $\chi^2$ analysis have shifted due to the type contamination.  One can attempt to fix this bias with simulations about the best fit value but one can use SMC ABC to reproduce the full bias-correct contours.  The ABC contours are 42\% larger in area than the $\chi^2$ uncontaminated contours, but
cover essentially the same area as the original ABC contours from the uncontaminated
sample.   If contamination is properly modeled the ABC method is robust against
these effects that can only be applied on a population basis rather than as a
per-object correction.

It is worthwhile to note that while the division between statistical and systematic errors is often loosely used to make a distinction between uncertainties that will decrease with more data of the same form versus uncertainties that will not decrease with larger sample sizes, the benefit of a forward-modeling framework is that they can be treated consistently and simultaneously.
To create a simulation model one is forced to make choices regarding the distributions of all statistical and systematic uncertainties through either analytic or empirical methods.  
Systematic errors come in at least three flavors: (1) effects that we know and understand and have a reasonable understanding of the relevant input distribution; (2) effects we qualitatively understand, but for which we do not have a good input prior distribution: e.g., $R_V$ values in host galaxies.  We can compute the effect on a supernova lightcurve, but we are relatively uncertain about the correct distribution of $R_V$ in galaxies in the Universe; (3) effects that we lose sleep over but that we have so little understanding of that we cannot model their effects at all, although we may have some purely empirical guidance: spectroscopic selection biases; evolving metallicity content of stars over the last 8 billion years.  Systematic errors of type 1 are easy to include in ABC.  One can use ABC to examine the effects on the posterior distribution from different choices of distributions for systematic errors of type 2. One may be able to include empirical distributions for systematic errors of type 3. Otherwise ABC can not tell you something about these systematic errors unless they are treated as model parameters.   
Forward modeling with an SMC ABC approach provides a powerful way to fully incorporate all available knowledge and ignorance.






\section{Future Work}
\label{sec:future_work}

We presented here a proof of concept for an SMC ABC method to infer model parameters based on SN~Ia measurements.  To fully deploy this method will require an incorporation of all data sets and modeling relative systematics between the surveys, e.g., relative calibration.  This is tractable, if somewhat tedious, and has been done with varying degrees of completeness already in the literature.  Extending this approach to explorations of time-variable dark energy is a simple matter of implementing at different generating model for luminosity distance as a function of redshift.

For future photometric-focused surveys, we would explore more fully the non-Gaussianity of photometric redshifts as derived from calibration samples.   The probability distributions for these photometric redshifts will be strongly affected by evolution of the contamination fraction of non-SN~Ia with redshift.  Once that is phrased as part of the generating model, ABC will incorporate such uncertainties on the same basis as all of the other cosmological and astrophysical parameters.

The ABC+SNANA framework is a very suitable vehicle for testing the effects of different lightcurve fitters on the derived cosmological parameters.  ABC will help efficiently determine what different parameter choices in the fitters should be explored.

But the real long-term goal would be to apply the summary statistic comparison at the individual lightcurve level.  This could significantly reduce the computing time.  The analysis presented in this work with $\sim$100 supernovae and 1200 particles required $\sim$600 CPU-hours.  We estimate that a realistic problem with a sample of $10^4$ supernovae could be done on $O(10)$ CPU-years, which is within reasonable computing resources.  Applying the summary statistic comparison at the individual light curve level rather than in Hubble diagram space bypasses fitting the simulated light curves which currently requires most ($\sim$90\%) of the computing time. 

Comparing the simulated and observed data at the individual lightcurve level would also be the cleanest framework to explore agreement and evolution of systematics.  The only ``training'' would be in the generation of the templates that the SN~Ia are derived from in the first place.  The cosmological distance and supernova property comparison would be finally integrated in one direct comparison.

\section{Conclusions}
\label{sec:Conclusion}

We have introduced and demonstrated the use of Approximate Bayesian Computation techniques to address the requirements for analyzing near-future SN~Ia cosmological data sets.  ABC presents a consistent and efficient approach to explore multi-dimensional non-Gaussian parameter distributions with full incorporation of systematic uncertainties.

\begin{itemize}
\item Forward modeling is often the only way to correctly incorporate the full range of statistical and systematic uncertainties in some of the big astronomy questions being addressed today.
\item Calculation of likelihood functions for evaluation in a traditional Markov Chain Monte Carlo approach may not be analytically tractable.
\item ABC allows for a simultaneous exploration of parameter space and tolerance to create credible regions for physical parameters of interest without the need to construct an explicit likelihood function.
\item Sequential Monte Carlo ABC offers an efficient way to explore the full parameter space of all important input parameters and model effects.
\item The use of a summary statistic focuses attention directly on the ability to discriminate model parameter values in the relevant space of observed values.
\end{itemize}

We encourage scientists facing similar problems to consider the use of ABC techniques to increase their incisive power to explore the complicated parameter spaces that are surrounding the key questions in astrophysics and cosmology today.

\section{Acknowledgements}

We thank the referee for helpful comments and A.W. and M.W.-V. thank Saurabh Jha for insightful discussions.  This research was supported in part by NSF DMS-1106956.  A.W. and M.W.-V. were
supported in part by NSF AST-1028162. A.W. additionally acknowledges support from PITT PACC and the Zaccheus Daniel Foundation.


\appendix
\section{Non-Parametrically Smoothing the Simulated and Observed Data}
\label{appendix:smooth}

To perform a non-parametric smooth we use a robust locally weighted regression (loess)\citep{Cleveland79}.  This routine smooths the data by iteratively fitting a local d-order polynomial to the data using a tricube weighting function.  We use a quadratic polynomial and, for the observed data, add an additional weight according to the uncertainty in $\mu$ given by Eq. \ref{eq:sigmamu}.  

We choose the size of the window to locally smooth over by minimizing the risk or the sum of the variance and bias squared.  We estimate the risk using the leave-one-out cross validation score 
\begin{equation}
R(h) = \frac{1}{N}\sum_{i=0}^I (f(x_i) - f_{(-x_i)}(x_i))^2 
\end{equation}
where $f(x)$ is the smoothed function using a smoothing window given by $h$ and $f_{(-x_i)}$ is the smooth obtained leaving out the $i^{\rm th}$ data point (see, e.g., \citet{Wasserman06}).  The smoothing window goes from zero to one with zero being no smooth and one resulting in a line.  Using the SDSS data we find the minimum risk to yield a smoothing window of 0.52.  As estimating the risk is somewhat computationally intensive, we determine the smoothing window using the observed data and use the same window to smooth the simulated data in the ABC algorithm.

\bibliographystyle{astroads}
\bibliography{ABC}

\begin{thebibliography}{51}
\expandafter\ifx\csname natexlab\endcsname\relax\def\natexlab#1{#1}\fi
\expandafter\ifx\csname href\endcsname\relax
  \def\href#1#2{}\fi
\expandafter\ifx\csname urllinklabel\endcsname\relax
  \def\urllinklabel{[LINK]}\fi
\expandafter\ifx\csname adsurllinklabel\endcsname\relax
  \def\adsurllinklabel{[ADS]}\fi

\bibitem[{{Amanullah} {et~al.}(2010){Amanullah}, {Lidman}, {Rubin}, {Aldering},
  {Astier}, {Barbary}, {Burns}, {Conley}, {Dawson}, {Deustua}, {Doi}, {Fabbro},
  {Faccioli}, {Fakhouri}, {Folatelli}, {Fruchter}, {Furusawa}, {Garavini},
  {Goldhaber}, {Goobar}, {Groom}, {Hook}, {Howell}, {Kashikawa}, {Kim}, {Knop},
  {Kowalski}, {Linder}, {Meyers}, {Morokuma}, {Nobili}, {Nordin}, {Nugent},
  {{\"O}stman}, {Pain}, {Panagia}, {Perlmutter}, {Raux}, {Ruiz-Lapuente},
  {Spadafora}, {Strovink}, {Suzuki}, {Wang}, {Wood-Vasey}, {Yasuda}, \&
  {Supernova Cosmology Project}}]{Amanullah10}
{Amanullah}, R., {Lidman}, C., {Rubin}, D., {Aldering}, G., {Astier}, P.,
  {Barbary}, K., {Burns}, M.~S., {Conley}, A., {Dawson}, K.~S., {Deustua},
  S.~E., {Doi}, M., {Fabbro}, S., {Faccioli}, L., {Fakhouri}, H.~K.,
  {Folatelli}, G., {Fruchter}, A.~S., {Furusawa}, H., {Garavini}, G.,
  {Goldhaber}, G., {Goobar}, A., {Groom}, D.~E., {Hook}, I., {Howell}, D.~A.,
  {Kashikawa}, N., {Kim}, A.~G., {Knop}, R.~A., {Kowalski}, M., {Linder}, E.,
  {Meyers}, J., {Morokuma}, T., {Nobili}, S., {Nordin}, J., {Nugent}, P.~E.,
  {{\"O}stman}, L., {Pain}, R., {Panagia}, N., {Perlmutter}, S., {Raux}, J.,
  {Ruiz-Lapuente}, P., {Spadafora}, A.~L., {Strovink}, M., {Suzuki}, N.,
  {Wang}, L., {Wood-Vasey}, W.~M., {Yasuda}, N., \& {Supernova Cosmology
  Project}, T. 2010, \apj, 716, 712


\bibitem[{{Astier} {et~al.}(2006){Astier}, {Guy}, {Regnault}, {Pain},
  {Aubourg}, {Balam}, {Basa}, {Carlberg}, {Fabbro}, {Fouchez}, {Hook},
  {Howell}, {Lafoux}, {Neill}, {Palanque-Delabrouille}, {Perrett}, {Pritchet},
  {Rich}, {Sullivan}, {Taillet}, {Aldering}, {Antilogus}, {Arsenijevic},
  {Balland}, {Baumont}, {Bronder}, {Courtois}, {Ellis}, {Filiol}, {Gon{\c
  c}alves}, {Goobar}, {Guide}, {Hardin}, {Lusset}, {Lidman}, {McMahon},
  {Mouchet}, {Mourao}, {Perlmutter}, {Ripoche}, {Tao}, \& {Walton}}]{Astier06}
{Astier}, P., {Guy}, J., {Regnault}, N., {Pain}, R., {Aubourg}, E., {Balam},
  D., {Basa}, S., {Carlberg}, R.~G., {Fabbro}, S., {Fouchez}, D., {Hook},
  I.~M., {Howell}, D.~A., {Lafoux}, H., {Neill}, J.~D.,
  {Palanque-Delabrouille}, N., {Perrett}, K., {Pritchet}, C.~J., {Rich}, J.,
  {Sullivan}, M., {Taillet}, R., {Aldering}, G., {Antilogus}, P.,
  {Arsenijevic}, V., {Balland}, C., {Baumont}, S., {Bronder}, J., {Courtois},
  H., {Ellis}, R.~S., {Filiol}, M., {Gon{\c c}alves}, A.~C., {Goobar}, A.,
  {Guide}, D., {Hardin}, D., {Lusset}, V., {Lidman}, C., {McMahon}, R.,
  {Mouchet}, M., {Mourao}, A., {Perlmutter}, S., {Ripoche}, P., {Tao}, C., \&
  {Walton}, N. 2006, \aap, 447, 31


\bibitem[{{Barlow}(2003)}]{barlow03}
{Barlow}, R. 2003, ArXiv Physics e-prints


\bibitem[{{Barnes} {et~al.}(2011){Barnes}, {Filippi}, {Stumpf}, \&
  {Thorne}}]{Barnes11}
{Barnes}, C., {Filippi}, S., {Stumpf}, M.~P.~H., \& {Thorne}, T. 2011, ArXiv
  e-prints, 1106.6281


\bibitem[{Beaumont {et~al.}(2009)Beaumont, Cornuet, Marin, \&
  Robert}]{Beaumont09}
Beaumont, M.~A., Cornuet, J.-M., Marin, J.-M., \& Robert, C.~P. 2009,
  Biometrika, 96, 983


\bibitem[{Beaumont {et~al.}(2002)Beaumont, Zhang, \& Balding}]{Beaumont02}
Beaumont, M.~A., Zhang, W., \& Balding, D.~J. 2002, Genetics, 162, 2025


\bibitem[{{Blum} \& {Francois}(2010)}]{Blum08}
{Blum}, M.~G.~B. \& {Francois}, O. 2010, Statistics and Computing, 20, 63


\bibitem[{{Blum} {et~al.}(2012){Blum}, {Nunes}, {Prangle}, \&
  {Sisson}}]{Blum2012}
{Blum}, M.~G.~B., {Nunes}, M.~A., {Prangle}, D., \& {Sisson}, S.~A. 2012, ArXiv
  e-prints, 1202.3819


\bibitem[{{Cameron} \& {Pettitt}(2012)}]{Cameron12}
{Cameron}, E. \& {Pettitt}, A.~N. 2012, ArXiv e-prints, 1202.1426


\bibitem[{{Cardelli} {et~al.}(1989){Cardelli}, {Clayton}, \&
  {Mathis}}]{Cardelli89}
{Cardelli}, J.~A., {Clayton}, G.~C., \& {Mathis}, J.~S. 1989, \apj, 345, 245


\bibitem[{Cleveland(1979)}]{Cleveland79}
Cleveland, W.~S. 1979, Journal of the American Statistical Association, 74, 829
 \href{http://www.tandfonline.com/doi/abs/10.1080/01621459.1979.10481038}{\url%
linklabel}

\bibitem[{{Conley} {et~al.}(2011){Conley}, {Guy}, {Sullivan}, {Regnault},
  {Astier}, {Balland}, {Basa}, {Carlberg}, {Fouchez}, {Hardin}, {Hook},
  {Howell}, {Pain}, {Palanque-Delabrouille}, {Perrett}, {Pritchet}, {Rich},
  {Ruhlmann-Kleider}, {Balam}, {Baumont}, {Ellis}, {Fabbro}, {Fakhouri},
  {Fourmanoit}, {Gonz{\'a}lez-Gait{\'a}n}, {Graham}, {Hudson}, {Hsiao},
  {Kronborg}, {Lidman}, {Mourao}, {Neill}, {Perlmutter}, {Ripoche}, {Suzuki},
  \& {Walker}}]{Conley11}
{Conley}, A., {Guy}, J., {Sullivan}, M., {Regnault}, N., {Astier}, P.,
  {Balland}, C., {Basa}, S., {Carlberg}, R.~G., {Fouchez}, D., {Hardin}, D.,
  {Hook}, I.~M., {Howell}, D.~A., {Pain}, R., {Palanque-Delabrouille}, N.,
  {Perrett}, K.~M., {Pritchet}, C.~J., {Rich}, J., {Ruhlmann-Kleider}, V.,
  {Balam}, D., {Baumont}, S., {Ellis}, R.~S., {Fabbro}, S., {Fakhouri}, H.~K.,
  {Fourmanoit}, N., {Gonz{\'a}lez-Gait{\'a}n}, S., {Graham}, M.~L., {Hudson},
  M.~J., {Hsiao}, E., {Kronborg}, T., {Lidman}, C., {Mourao}, A.~M., {Neill},
  J.~D., {Perlmutter}, S., {Ripoche}, P., {Suzuki}, N., \& {Walker}, E.~S.
  2011, \apjs, 192, 1


\bibitem[{{Contreras} {et~al.}(2010){Contreras}, {Hamuy}, {Phillips},
  {Folatelli}, {Suntzeff}, {Persson}, {Stritzinger}, {Boldt}, {Gonz{\'a}lez},
  {Krzeminski}, {Morrell}, {Roth}, {Salgado}, {Jos{\'e} Maureira}, {Burns},
  {Freedman}, {Madore}, {Murphy}, {Wyatt}, {Li}, \& {Filippenko}}]{Contreras10}
{Contreras}, C., {Hamuy}, M., {Phillips}, M.~M., {Folatelli}, G., {Suntzeff},
  N.~B., {Persson}, S.~E., {Stritzinger}, M., {Boldt}, L., {Gonz{\'a}lez}, S.,
  {Krzeminski}, W., {Morrell}, N., {Roth}, M., {Salgado}, F., {Jos{\'e}
  Maureira}, M., {Burns}, C.~R., {Freedman}, W.~L., {Madore}, B.~F., {Murphy},
  D., {Wyatt}, P., {Li}, W., \& {Filippenko}, A.~V. 2010, \aj, 139, 519


\bibitem[{Cowles \& Carlin(1996)}]{Cowles96}
Cowles, M.~K. \& Carlin, B.~P. 1996, Journal of the American Statistical
  Association, 91, pp. 883
 \href{http://www.jstor.org/stable/2291683}{\urllinklabel}

\bibitem[{{Dilday} {et~al.}(2008){Dilday}, {Kessler}, {Frieman}, {Holtzman},
  {Marriner}, {Miknaitis}, {Nichol}, {Romani}, {Sako}, {Bassett}, {Becker},
  {Cinabro}, {DeJongh}, {Depoy}, {Doi}, {Garnavich}, {Hogan}, {Jha}, {Konishi},
  {Lampeitl}, {Marshall}, {McGinnis}, {Prieto}, {Riess}, {Richmond},
  {Schneider}, {Smith}, {Takanashi}, {Tokita}, {van der Heyden}, {Yasuda},
  {Zheng}, {Barentine}, {Brewington}, {Choi}, {Crotts}, {Dembicky}, {Harvanek},
  {Im}, {Ketzeback}, {Kleinman}, {Krzesi{\'n}ski}, {Long}, {Malanushenko},
  {Malanushenko}, {McMillan}, {Nitta}, {Pan}, {Saurage}, {Snedden}, {Watters},
  {Wheeler}, \& {York}}]{Dilday08}
{Dilday}, B., {Kessler}, R., {Frieman}, J.~A., {Holtzman}, J., {Marriner}, J.,
  {Miknaitis}, G., {Nichol}, R.~C., {Romani}, R., {Sako}, M., {Bassett}, B.,
  {Becker}, A., {Cinabro}, D., {DeJongh}, F., {Depoy}, D.~L., {Doi}, M.,
  {Garnavich}, P.~M., {Hogan}, C.~J., {Jha}, S., {Konishi}, K., {Lampeitl}, H.,
  {Marshall}, J.~L., {McGinnis}, D., {Prieto}, J.~L., {Riess}, A.~G.,
  {Richmond}, M.~W., {Schneider}, D.~P., {Smith}, M., {Takanashi}, N.,
  {Tokita}, K., {van der Heyden}, K., {Yasuda}, N., {Zheng}, C., {Barentine},
  J., {Brewington}, H., {Choi}, C., {Crotts}, A., {Dembicky}, J., {Harvanek},
  M., {Im}, M., {Ketzeback}, W., {Kleinman}, S.~J., {Krzesi{\'n}ski}, J.,
  {Long}, D.~C., {Malanushenko}, E., {Malanushenko}, V., {McMillan}, R.~J.,
  {Nitta}, A., {Pan}, K., {Saurage}, G., {Snedden}, S.~A., {Watters}, S.,
  {Wheeler}, J.~C., \& {York}, D. 2008, \apj, 682, 262


\bibitem[{{Eisenstein} {et~al.}(2005){Eisenstein}, {Zehavi}, {Hogg},
  {Scoccimarro}, {Blanton}, {Nichol}, {Scranton}, {Seo}, {Tegmark}, {Zheng},
  {Anderson}, {Annis}, {Bahcall}, {Brinkmann}, {Burles}, {Castander},
  {Connolly}, {Csabai}, {Doi}, {Fukugita}, {Frieman}, {Glazebrook}, {Gunn},
  {Hendry}, {Hennessy}, {Ivezi{\'c}}, {Kent}, {Knapp}, {Lin}, {Loh}, {Lupton},
  {Margon}, {McKay}, {Meiksin}, {Munn}, {Pope}, {Richmond}, {Schlegel},
  {Schneider}, {Shimasaku}, {Stoughton}, {Strauss}, {SubbaRao}, {Szalay},
  {Szapudi}, {Tucker}, {Yanny}, \& {York}}]{Eisenstein05}
{Eisenstein}, D.~J., {Zehavi}, I., {Hogg}, D.~W., {Scoccimarro}, R., {Blanton},
  M.~R., {Nichol}, R.~C., {Scranton}, R., {Seo}, H.-J., {Tegmark}, M., {Zheng},
  Z., {Anderson}, S.~F., {Annis}, J., {Bahcall}, N., {Brinkmann}, J., {Burles},
  S., {Castander}, F.~J., {Connolly}, A., {Csabai}, I., {Doi}, M., {Fukugita},
  M., {Frieman}, J.~A., {Glazebrook}, K., {Gunn}, J.~E., {Hendry}, J.~S.,
  {Hennessy}, G., {Ivezi{\'c}}, Z., {Kent}, S., {Knapp}, G.~R., {Lin}, H.,
  {Loh}, Y.-S., {Lupton}, R.~H., {Margon}, B., {McKay}, T.~A., {Meiksin}, A.,
  {Munn}, J.~A., {Pope}, A., {Richmond}, M.~W., {Schlegel}, D., {Schneider},
  D.~P., {Shimasaku}, K., {Stoughton}, C., {Strauss}, M.~A., {SubbaRao}, M.,
  {Szalay}, A.~S., {Szapudi}, I., {Tucker}, D.~L., {Yanny}, B., \& {York},
  D.~G. 2005, \apj, 633, 560


\bibitem[{{Fearnhead} \& {Prangle}(2012)}]{Fearnhead2012}
{Fearnhead}, P. \& {Prangle}, D. 2012, Journal Of The Royal Statistical Society
  Series B, 74


\bibitem[{{Filippi} {et~al.}(2011){Filippi}, {Barnes}, \& {Stumpf}}]{Filippi11}
{Filippi}, S., {Barnes}, C., \& {Stumpf}, M.~P.~H. 2011, ArXiv e-prints,
  1106.6280


\bibitem[{{Frieman} {et~al.}(2008){Frieman}, {Bassett}, {Becker}, {Choi},
  {Cinabro}, {DeJongh}, {Depoy}, {Dilday}, {Doi}, {Garnavich}, {Hogan},
  {Holtzman}, {Im}, {Jha}, {Kessler}, {Konishi}, {Lampeitl}, {Marriner},
  {Marshall}, {McGinnis}, {Miknaitis}, {Nichol}, {Prieto}, {Riess}, {Richmond},
  {Romani}, {Sako}, {Schneider}, {Smith}, {Takanashi}, {Tokita}, {van der
  Heyden}, {Yasuda}, {Zheng}, {Adelman-McCarthy}, {Annis}, {Assef},
  {Barentine}, {Bender}, {Blandford}, {Boroski}, {Bremer}, {Brewington},
  {Collins}, {Crotts}, {Dembicky}, {Eastman}, {Edge}, {Edmondson}, {Elson},
  {Eyler}, {Filippenko}, {Foley}, {Frank}, {Goobar}, {Gueth}, {Gunn},
  {Harvanek}, {Hopp}, {Ihara}, {Ivezi{\'c}}, {Kahn}, {Kaplan}, {Kent},
  {Ketzeback}, {Kleinman}, {Kollatschny}, {Kron}, {Krzesi{\'n}ski}, {Lamenti},
  {Leloudas}, {Lin}, {Long}, {Lucey}, {Lupton}, {Malanushenko}, {Malanushenko},
  {McMillan}, {Mendez}, {Morgan}, {Morokuma}, {Nitta}, {Ostman}, {Pan},
  {Rockosi}, {Romer}, {Ruiz-Lapuente}, {Saurage}, {Schlesinger}, {Snedden},
  {Sollerman}, {Stoughton}, {Stritzinger}, {Subba Rao}, {Tucker}, {Vaisanen},
  {Watson}, {Watters}, {Wheeler}, {Yanny}, \& {York}}]{Frieman08}
{Frieman}, J.~A., {Bassett}, B., {Becker}, A., {Choi}, C., {Cinabro}, D.,
  {DeJongh}, F., {Depoy}, D.~L., {Dilday}, B., {Doi}, M., {Garnavich}, P.~M.,
  {Hogan}, C.~J., {Holtzman}, J., {Im}, M., {Jha}, S., {Kessler}, R.,
  {Konishi}, K., {Lampeitl}, H., {Marriner}, J., {Marshall}, J.~L., {McGinnis},
  D., {Miknaitis}, G., {Nichol}, R.~C., {Prieto}, J.~L., {Riess}, A.~G.,
  {Richmond}, M.~W., {Romani}, R., {Sako}, M., {Schneider}, D.~P., {Smith}, M.,
  {Takanashi}, N., {Tokita}, K., {van der Heyden}, K., {Yasuda}, N., {Zheng},
  C., {Adelman-McCarthy}, J., {Annis}, J., {Assef}, R.~J., {Barentine}, J.,
  {Bender}, R., {Blandford}, R.~D., {Boroski}, W.~N., {Bremer}, M.,
  {Brewington}, H., {Collins}, C.~A., {Crotts}, A., {Dembicky}, J., {Eastman},
  J., {Edge}, A., {Edmondson}, E., {Elson}, E., {Eyler}, M.~E., {Filippenko},
  A.~V., {Foley}, R.~J., {Frank}, S., {Goobar}, A., {Gueth}, T., {Gunn}, J.~E.,
  {Harvanek}, M., {Hopp}, U., {Ihara}, Y., {Ivezi{\'c}}, {\v Z}., {Kahn}, S.,
  {Kaplan}, J., {Kent}, S., {Ketzeback}, W., {Kleinman}, S.~J., {Kollatschny},
  W., {Kron}, R.~G., {Krzesi{\'n}ski}, J., {Lamenti}, D., {Leloudas}, G.,
  {Lin}, H., {Long}, D.~C., {Lucey}, J., {Lupton}, R.~H., {Malanushenko}, E.,
  {Malanushenko}, V., {McMillan}, R.~J., {Mendez}, J., {Morgan}, C.~W.,
  {Morokuma}, T., {Nitta}, A., {Ostman}, L., {Pan}, K., {Rockosi}, C.~M.,
  {Romer}, A.~K., {Ruiz-Lapuente}, P., {Saurage}, G., {Schlesinger}, K.,
  {Snedden}, S.~A., {Sollerman}, J., {Stoughton}, C., {Stritzinger}, M., {Subba
  Rao}, M., {Tucker}, D., {Vaisanen}, P., {Watson}, L.~C., {Watters}, S.,
  {Wheeler}, J.~C., {Yanny}, B., \& {York}, D. 2008, \aj, 135, 338


\bibitem[{Fu \& Li(1997)}]{Fu97}
Fu, Y.~X. \& Li, W.~H. 1997, Molecular Biology and Evolution, 14, 195


\bibitem[{{Ganeshalingam} {et~al.}(2010){Ganeshalingam}, {Li}, {Filippenko},
  {Anderson}, {Foster}, {Gates}, {Griffith}, {Grigsby}, {Joubert}, {Leja},
  {Lowe}, {Macomber}, {Pritchard}, {Thrasher}, \& {Winslow}}]{Gane10}
{Ganeshalingam}, M., {Li}, W., {Filippenko}, A.~V., {Anderson}, C., {Foster},
  G., {Gates}, E.~L., {Griffith}, C.~V., {Grigsby}, B.~J., {Joubert}, N.,
  {Leja}, J., {Lowe}, T.~B., {Macomber}, B., {Pritchard}, T., {Thrasher}, P.,
  \& {Winslow}, D. 2010, \apjs, 190, 418


\bibitem[{{Hicken} {et~al.}(2012){Hicken}, {Challis}, {Kirshner}, {Rest},
  {Cramer}, {Wood-Vasey}, {Bakos}, {Berlind}, {Brown}, {Caldwell}, {Calkins},
  {Currie}, {de Kleer}, {Esquerdo}, {Everett}, {Falco}, {Fernandez},
  {Friedman}, {Groner}, {Hartman}, {Holman}, {Hutchins}, {Keys}, {Kipping},
  {Latham}, {Marion}, {Narayan}, {Pahre}, {Pal}, {Peters}, {Perumpilly},
  {Ripman}, {Sipocz}, {Szentgyorgyi}, {Tang}, {Torres}, {Vaz}, {Wolk}, \&
  {Zezas}}]{Hicken12}
{Hicken}, M., {Challis}, P., {Kirshner}, R.~P., {Rest}, A., {Cramer}, C.~E.,
  {Wood-Vasey}, W.~M., {Bakos}, G., {Berlind}, P., {Brown}, W.~R., {Caldwell},
  N., {Calkins}, M., {Currie}, T., {de Kleer}, K., {Esquerdo}, G., {Everett},
  M., {Falco}, E., {Fernandez}, J., {Friedman}, A.~S., {Groner}, T., {Hartman},
  J., {Holman}, M.~J., {Hutchins}, R., {Keys}, S., {Kipping}, D., {Latham}, D.,
  {Marion}, G.~H., {Narayan}, G., {Pahre}, M., {Pal}, A., {Peters}, W.,
  {Perumpilly}, G., {Ripman}, B., {Sipocz}, B., {Szentgyorgyi}, A., {Tang}, S.,
  {Torres}, M.~A.~P., {Vaz}, A., {Wolk}, S., \& {Zezas}, A. 2012, \apjs, 200,
  12


\bibitem[{{Hicken} {et~al.}(2009){Hicken}, {Wood-Vasey}, {Blondin}, {Challis},
  {Jha}, {Kelly}, {Rest}, \& {Kirshner}}]{Hicken09b}
{Hicken}, M., {Wood-Vasey}, W.~M., {Blondin}, S., {Challis}, P., {Jha}, S.,
  {Kelly}, P.~L., {Rest}, A., \& {Kirshner}, R.~P. 2009, \apj, 700, 1097


\bibitem[{{Holtzman} {et~al.}(2008){Holtzman}, {Marriner}, {Kessler}, {Sako},
  {Dilday}, {Frieman}, {Schneider}, {Bassett}, {Becker}, {Cinabro}, {DeJongh},
  {Depoy}, {Doi}, {Garnavich}, {Hogan}, {Jha}, {Konishi}, {Lampeitl},
  {Marshall}, {McGinnis}, {Miknaitis}, {Nichol}, {Prieto}, {Riess}, {Richmond},
  {Romani}, {Smith}, {Takanashi}, {Tokita}, {van der Heyden}, {Yasuda}, \&
  {Zheng}}]{Holtzman08}
{Holtzman}, J.~A., {Marriner}, J., {Kessler}, R., {Sako}, M., {Dilday}, B.,
  {Frieman}, J.~A., {Schneider}, D.~P., {Bassett}, B., {Becker}, A., {Cinabro},
  D., {DeJongh}, F., {Depoy}, D.~L., {Doi}, M., {Garnavich}, P.~M., {Hogan},
  C.~J., {Jha}, S., {Konishi}, K., {Lampeitl}, H., {Marshall}, J.~L.,
  {McGinnis}, D., {Miknaitis}, G., {Nichol}, R.~C., {Prieto}, J.~L., {Riess},
  A.~G., {Richmond}, M.~W., {Romani}, R., {Smith}, M., {Takanashi}, N.,
  {Tokita}, K., {van der Heyden}, K., {Yasuda}, N., \& {Zheng}, C. 2008, \aj,
  136, 2306


\bibitem[{{Hsiao} {et~al.}(2007){Hsiao}, {Conley}, {Howell}, {Sullivan},
  {Pritchet}, {Carlberg}, {Nugent}, \& {Phillips}}]{Hsiao07}
{Hsiao}, E.~Y., {Conley}, A., {Howell}, D.~A., {Sullivan}, M., {Pritchet},
  C.~J., {Carlberg}, R.~G., {Nugent}, P.~E., \& {Phillips}, M.~M. 2007, \apj,
  663, 1187


\bibitem[{{Jha} {et~al.}(2007){Jha}, {Riess}, \& {Kirshner}}]{Jha07}
{Jha}, S., {Riess}, A.~G., \& {Kirshner}, R.~P. 2007, \apj, 659, 122


\bibitem[{{Kessler} {et~al.}(2009{\natexlab{a}}){Kessler}, {Becker}, {Cinabro},
  {Vanderplas}, {Frieman}, {Marriner}, {Davis}, {Dilday}, {Holtzman}, {Jha},
  {Lampeitl}, {Sako}, {Smith}, {Zheng}, {Nichol}, {Bassett}, {Bender}, {Depoy},
  {Doi}, {Elson}, {Filippenko}, {Foley}, {Garnavich}, {Hopp}, {Ihara},
  {Ketzeback}, {Kollatschny}, {Konishi}, {Marshall}, {McMillan}, {Miknaitis},
  {Morokuma}, {M{\"o}rtsell}, {Pan}, {Prieto}, {Richmond}, {Riess}, {Romani},
  {Schneider}, {Sollerman}, {Takanashi}, {Tokita}, {van der Heyden}, {Wheeler},
  {Yasuda}, \& {York}}]{Kessler09b}
{Kessler}, R., {Becker}, A.~C., {Cinabro}, D., {Vanderplas}, J., {Frieman},
  J.~A., {Marriner}, J., {Davis}, T.~M., {Dilday}, B., {Holtzman}, J., {Jha},
  S.~W., {Lampeitl}, H., {Sako}, M., {Smith}, M., {Zheng}, C., {Nichol}, R.~C.,
  {Bassett}, B., {Bender}, R., {Depoy}, D.~L., {Doi}, M., {Elson}, E.,
  {Filippenko}, A.~V., {Foley}, R.~J., {Garnavich}, P.~M., {Hopp}, U., {Ihara},
  Y., {Ketzeback}, W., {Kollatschny}, W., {Konishi}, K., {Marshall}, J.~L.,
  {McMillan}, R.~J., {Miknaitis}, G., {Morokuma}, T., {M{\"o}rtsell}, E.,
  {Pan}, K., {Prieto}, J.~L., {Richmond}, M.~W., {Riess}, A.~G., {Romani}, R.,
  {Schneider}, D.~P., {Sollerman}, J., {Takanashi}, N., {Tokita}, K., {van der
  Heyden}, K., {Wheeler}, J.~C., {Yasuda}, N., \& {York}, D.
  2009{\natexlab{a}}, \apjs, 185, 32


\bibitem[{{Kessler} {et~al.}(2009{\natexlab{b}}){Kessler}, {Bernstein},
  {Cinabro}, {Dilday}, {Frieman}, {Jha}, {Kuhlmann}, {Miknaitis}, {Sako},
  {Taylor}, \& {Vanderplas}}]{Kessler09a}
{Kessler}, R., {Bernstein}, J.~P., {Cinabro}, D., {Dilday}, B., {Frieman},
  J.~A., {Jha}, S., {Kuhlmann}, S., {Miknaitis}, G., {Sako}, M., {Taylor}, M.,
  \& {Vanderplas}, J. 2009{\natexlab{b}}, \pasp, 121, 1028


\bibitem[{{Knop} {et~al.}(2003){Knop}, {Aldering}, {Amanullah}, {Astier},
  {Blanc}, {Burns}, {Conley}, {Deustua}, {Doi}, {Ellis}, {Fabbro}, {Folatelli},
  {Fruchter}, {Garavini}, {Garmond}, {Garton}, {Gibbons}, {Goldhaber},
  {Goobar}, {Groom}, {Hardin}, {Hook}, {Howell}, {Kim}, {Lee}, {Lidman},
  {Mendez}, {Nobili}, {Nugent}, {Pain}, {Panagia}, {Pennypacker}, {Perlmutter},
  {Quimby}, {Raux}, {Regnault}, {Ruiz-Lapuente}, {Sainton}, {Schaefer},
  {Schahmaneche}, {Smith}, {Spadafora}, {Stanishev}, {Sullivan}, {Walton},
  {Wang}, {Wood-Vasey}, \& {Yasuda}}]{Knop03}
{Knop}, R.~A., {Aldering}, G., {Amanullah}, R., {Astier}, P., {Blanc}, G.,
  {Burns}, M.~S., {Conley}, A., {Deustua}, S.~E., {Doi}, M., {Ellis}, R.,
  {Fabbro}, S., {Folatelli}, G., {Fruchter}, A.~S., {Garavini}, G., {Garmond},
  S., {Garton}, K., {Gibbons}, R., {Goldhaber}, G., {Goobar}, A., {Groom},
  D.~E., {Hardin}, D., {Hook}, I., {Howell}, D.~A., {Kim}, A.~G., {Lee}, B.~C.,
  {Lidman}, C., {Mendez}, J., {Nobili}, S., {Nugent}, P.~E., {Pain}, R.,
  {Panagia}, N., {Pennypacker}, C.~R., {Perlmutter}, S., {Quimby}, R., {Raux},
  J., {Regnault}, N., {Ruiz-Lapuente}, P., {Sainton}, G., {Schaefer}, B.,
  {Schahmaneche}, K., {Smith}, E., {Spadafora}, A.~L., {Stanishev}, V.,
  {Sullivan}, M., {Walton}, N.~A., {Wang}, L., {Wood-Vasey}, W.~M., \&
  {Yasuda}, N. 2003, \apj, 598, 102


\bibitem[{{Komatsu} {et~al.}(2009){Komatsu}, {Dunkley}, {Nolta}, {Bennett},
  {Gold}, {Hinshaw}, {Jarosik}, {Larson}, {Limon}, {Page}, {Spergel},
  {Halpern}, {Hill}, {Kogut}, {Meyer}, {Tucker}, {Weiland}, {Wollack}, \&
  {Wright}}]{Komatsu09}
{Komatsu}, E., {Dunkley}, J., {Nolta}, M.~R., {Bennett}, C.~L., {Gold}, B.,
  {Hinshaw}, G., {Jarosik}, N., {Larson}, D., {Limon}, M., {Page}, L.,
  {Spergel}, D.~N., {Halpern}, M., {Hill}, R.~S., {Kogut}, A., {Meyer}, S.~S.,
  {Tucker}, G.~S., {Weiland}, J.~L., {Wollack}, E., \& {Wright}, E.~L. 2009,
  \apjs, 180, 330


\bibitem[{{Kowalski} {et~al.}(2008){Kowalski}, {Rubin}, {Aldering},
  {Agostinho}, {Amadon}, {Amanullah}, {Balland}, {Barbary}, {Blanc}, {Challis},
  {Conley}, {Connolly}, {Covarrubias}, {Dawson}, {Deustua}, {Ellis}, {Fabbro},
  {Fadeyev}, {Fan}, {Farris}, {Folatelli}, {Frye}, {Garavini}, {Gates},
  {Germany}, {Goldhaber}, {Goldman}, {Goobar}, {Groom}, {Haissinski}, {Hardin},
  {Hook}, {Kent}, {Kim}, {Knop}, {Lidman}, {Linder}, {Mendez}, {Meyers},
  {Miller}, {Moniez}, {Mour{\~a}o}, {Newberg}, {Nobili}, {Nugent}, {Pain},
  {Perdereau}, {Perlmutter}, {Phillips}, {Prasad}, {Quimby}, {Regnault},
  {Rich}, {Rubenstein}, {Ruiz-Lapuente}, {Santos}, {Schaefer}, {Schommer},
  {Smith}, {Soderberg}, {Spadafora}, {Strolger}, {Strovink}, {Suntzeff},
  {Suzuki}, {Thomas}, {Walton}, {Wang}, {Wood-Vasey}, \& {Yun}}]{Kowalski08}
{Kowalski}, M., {Rubin}, D., {Aldering}, G., {Agostinho}, R.~J., {Amadon}, A.,
  {Amanullah}, R., {Balland}, C., {Barbary}, K., {Blanc}, G., {Challis}, P.~J.,
  {Conley}, A., {Connolly}, N.~V., {Covarrubias}, R., {Dawson}, K.~S.,
  {Deustua}, S.~E., {Ellis}, R., {Fabbro}, S., {Fadeyev}, V., {Fan}, X.,
  {Farris}, B., {Folatelli}, G., {Frye}, B.~L., {Garavini}, G., {Gates}, E.~L.,
  {Germany}, L., {Goldhaber}, G., {Goldman}, B., {Goobar}, A., {Groom}, D.~E.,
  {Haissinski}, J., {Hardin}, D., {Hook}, I., {Kent}, S., {Kim}, A.~G., {Knop},
  R.~A., {Lidman}, C., {Linder}, E.~V., {Mendez}, J., {Meyers}, J., {Miller},
  G.~J., {Moniez}, M., {Mour{\~a}o}, A.~M., {Newberg}, H., {Nobili}, S.,
  {Nugent}, P.~E., {Pain}, R., {Perdereau}, O., {Perlmutter}, S., {Phillips},
  M.~M., {Prasad}, V., {Quimby}, R., {Regnault}, N., {Rich}, J., {Rubenstein},
  E.~P., {Ruiz-Lapuente}, P., {Santos}, F.~D., {Schaefer}, B.~E., {Schommer},
  R.~A., {Smith}, R.~C., {Soderberg}, A.~M., {Spadafora}, A.~L., {Strolger},
  L.-G., {Strovink}, M., {Suntzeff}, N.~B., {Suzuki}, N., {Thomas}, R.~C.,
  {Walton}, N.~A., {Wang}, L., {Wood-Vasey}, W.~M., \& {Yun}, J.~L. 2008, \apj,
  686, 749


\bibitem[{{Lampeitl} {et~al.}(2010){Lampeitl}, {Nichol}, {Seo}, {Giannantonio},
  {Shapiro}, {Bassett}, {Percival}, {Davis}, {Dilday}, {Frieman}, {Garnavich},
  {Sako}, {Smith}, {Sollerman}, {Becker}, {Cinabro}, {Filippenko}, {Foley},
  {Hogan}, {Holtzman}, {Jha}, {Konishi}, {Marriner}, {Richmond}, {Riess},
  {Schneider}, {Stritzinger}, {van der Heyden}, {Vanderplas}, {Wheeler}, \&
  {Zheng}}]{Lampeitl10}
{Lampeitl}, H., {Nichol}, R.~C., {Seo}, H.-J., {Giannantonio}, T., {Shapiro},
  C., {Bassett}, B., {Percival}, W.~J., {Davis}, T.~M., {Dilday}, B.,
  {Frieman}, J., {Garnavich}, P., {Sako}, M., {Smith}, M., {Sollerman}, J.,
  {Becker}, A.~C., {Cinabro}, D., {Filippenko}, A.~V., {Foley}, R.~J., {Hogan},
  C.~J., {Holtzman}, J.~A., {Jha}, S.~W., {Konishi}, K., {Marriner}, J.,
  {Richmond}, M.~W., {Riess}, A.~G., {Schneider}, D.~P., {Stritzinger}, M.,
  {van der Heyden}, K.~J., {Vanderplas}, J.~T., {Wheeler}, J.~C., \& {Zheng},
  C. 2010, \mnras, 401, 2331


\bibitem[{{Law} {et~al.}(2009){Law}, {Kulkarni}, {Dekany}, {Ofek}, {Quimby},
  {Nugent}, {Surace}, {Grillmair}, {Bloom}, {Kasliwal}, {Bildsten}, {Brown},
  {Cenko}, {Ciardi}, {Croner}, {Djorgovski}, {van Eyken}, {Filippenko}, {Fox},
  {Gal-Yam}, {Hale}, {Hamam}, {Helou}, {Henning}, {Howell}, {Jacobsen},
  {Laher}, {Mattingly}, {McKenna}, {Pickles}, {Poznanski}, {Rahmer}, {Rau},
  {Rosing}, {Shara}, {Smith}, {Starr}, {Sullivan}, {Velur}, {Walters}, \&
  {Zolkower}}]{law09}
{Law}, N.~M., {Kulkarni}, S.~R., {Dekany}, R.~G., {Ofek}, E.~O., {Quimby},
  R.~M., {Nugent}, P.~E., {Surace}, J., {Grillmair}, C.~C., {Bloom}, J.~S.,
  {Kasliwal}, M.~M., {Bildsten}, L., {Brown}, T., {Cenko}, S.~B., {Ciardi}, D.,
  {Croner}, E., {Djorgovski}, S.~G., {van Eyken}, J., {Filippenko}, A.~V.,
  {Fox}, D.~B., {Gal-Yam}, A., {Hale}, D., {Hamam}, N., {Helou}, G., {Henning},
  J., {Howell}, D.~A., {Jacobsen}, J., {Laher}, R., {Mattingly}, S., {McKenna},
  D., {Pickles}, A., {Poznanski}, D., {Rahmer}, G., {Rau}, A., {Rosing}, W.,
  {Shara}, M., {Smith}, R., {Starr}, D., {Sullivan}, M., {Velur}, V.,
  {Walters}, R., \& {Zolkower}, J. 2009, \pasp, 121, 1395


\bibitem[{{LSST Science Collaborations} {et~al.}(2009){LSST Science
  Collaborations}, {Abell}, {Allison}, {Anderson}, {Andrew}, {Angel}, {Armus},
  {Arnett}, {Asztalos}, {Axelrod}, \& et~al.}]{LSSTscienceBook}
{LSST Science Collaborations}, {Abell}, P.~A., {Allison}, J., {Anderson},
  S.~F., {Andrew}, J.~R., {Angel}, J.~R.~P., {Armus}, L., {Arnett}, D.,
  {Asztalos}, S.~J., {Axelrod}, T.~S., \& et~al. 2009, ArXiv e-prints,
  0912.0201


\bibitem[{{Mandel} {et~al.}(2011){Mandel}, {Narayan}, \& {Kirshner}}]{Mandel11}
{Mandel}, K.~S., {Narayan}, G., \& {Kirshner}, R.~P. 2011, \apj, 731, 120


\bibitem[{{Marin} {et~al.}(2011){Marin}, {Pudlo}, {Robert}, \&
  {Ryder}}]{Marin11}
{Marin}, J.-M., {Pudlo}, P., {Robert}, C.~P., \& {Ryder}, R. 2011, ArXiv
  e-prints, 1101.0955


\bibitem[{{Marjoram} {et~al.}(2003){Marjoram}, {Molitor}, {Plagnol}, \&
  {Tavar{\'e}}}]{Marjoram03}
{Marjoram}, P., {Molitor}, J., {Plagnol}, V., \& {Tavar{\'e}}, S. 2003,
  Proceedings of the National Academy of Science, 1001, 15324


\bibitem[{{Miknaitis} {et~al.}(2007){Miknaitis}, {Pignata}, {Rest},
  {Wood-Vasey}, {Blondin}, {Challis}, {Smith}, {Stubbs}, {Suntzeff}, {Foley},
  {Matheson}, {Tonry}, {Aguilera}, {Blackman}, {Becker}, {Clocchiatti},
  {Covarrubias}, {Davis}, {Filippenko}, {Garg}, {Garnavich}, {Hicken}, {Jha},
  {Krisciunas}, {Kirshner}, {Leibundgut}, {Li}, {Miceli}, {Narayan}, {Prieto},
  {Riess}, {Salvo}, {Schmidt}, {Sollerman}, {Spyromilio}, \&
  {Zenteno}}]{Miknaitis07}
{Miknaitis}, G., {Pignata}, G., {Rest}, A., {Wood-Vasey}, W.~M., {Blondin}, S.,
  {Challis}, P., {Smith}, R.~C., {Stubbs}, C.~W., {Suntzeff}, N.~B., {Foley},
  R.~J., {Matheson}, T., {Tonry}, J.~L., {Aguilera}, C., {Blackman}, J.~W.,
  {Becker}, A.~C., {Clocchiatti}, A., {Covarrubias}, R., {Davis}, T.~M.,
  {Filippenko}, A.~V., {Garg}, A., {Garnavich}, P.~M., {Hicken}, M., {Jha}, S.,
  {Krisciunas}, K., {Kirshner}, R.~P., {Leibundgut}, B., {Li}, W., {Miceli},
  A., {Narayan}, G., {Prieto}, J.~L., {Riess}, A.~G., {Salvo}, M.~E.,
  {Schmidt}, B.~P., {Sollerman}, J., {Spyromilio}, J., \& {Zenteno}, A. 2007,
  \apj, 666, 674


\bibitem[{Moral {et~al.}(2006)Moral, Doucet, \& Jasra}]{Moral06}
Moral, P.~D., Doucet, A., \& Jasra, A. 2006, Journal Of The Royal Statistical
  Society Series B, 68, 411
 \href{http://econpapers.repec.org/RePEc:bla:jorssb:v:68:y:2006:i:3:p:411-436}%
{\urllinklabel}

\bibitem[{{Perlmutter} {et~al.}(1999){Perlmutter}, {Aldering}, {Goldhaber},
  {Knop}, {Nugent}, {Castro}, {Deustua}, {Fabbro}, {Goobar}, {Groom}, {Hook},
  {Kim}, {Kim}, {Lee}, {Nunes}, {Pain}, {Pennypacker}, {Quimby}, {Lidman},
  {Ellis}, {Irwin}, {McMahon}, {Ruiz-Lapuente}, {Walton}, {Schaefer}, {Boyle},
  {Filippenko}, {Matheson}, {Fruchter}, {Panagia}, {Newberg}, {Couch}, \& {The
  Supernova Cosmology Project}}]{Perlmutter99}
{Perlmutter}, S., {Aldering}, G., {Goldhaber}, G., {Knop}, R.~A., {Nugent}, P.,
  {Castro}, P.~G., {Deustua}, S., {Fabbro}, S., {Goobar}, A., {Groom}, D.~E.,
  {Hook}, I.~M., {Kim}, A.~G., {Kim}, M.~Y., {Lee}, J.~C., {Nunes}, N.~J.,
  {Pain}, R., {Pennypacker}, C.~R., {Quimby}, R., {Lidman}, C., {Ellis}, R.~S.,
  {Irwin}, M., {McMahon}, R.~G., {Ruiz-Lapuente}, P., {Walton}, N., {Schaefer},
  B., {Boyle}, B.~J., {Filippenko}, A.~V., {Matheson}, T., {Fruchter}, A.~S.,
  {Panagia}, N., {Newberg}, H.~J.~M., {Couch}, W.~J., \& {The Supernova
  Cosmology Project}. 1999, \apj, 517, 565


\bibitem[{Pritchard {et~al.}(1999)Pritchard, Seielstad, Perez-Lezaun, \&
  Feldman}]{Pritchard99}
Pritchard, J.~K., Seielstad, M.~T., Perez-Lezaun, A., \& Feldman, M.~W. 1999,
  Molecular Biology and Evolution, 16, 1791


\bibitem[{{Riess} {et~al.}(1998){Riess}, {Filippenko}, {Challis},
  {Clocchiatti}, {Diercks}, {Garnavich}, {Gilliland}, {Hogan}, {Jha},
  {Kirshner}, {Leibundgut}, {Phillips}, {Reiss}, {Schmidt}, {Schommer},
  {Smith}, {Spyromilio}, {Stubbs}, {Suntzeff}, \& {Tonry}}]{Riess98}
{Riess}, A.~G., {Filippenko}, A.~V., {Challis}, P., {Clocchiatti}, A.,
  {Diercks}, A., {Garnavich}, P.~M., {Gilliland}, R.~L., {Hogan}, C.~J., {Jha},
  S., {Kirshner}, R.~P., {Leibundgut}, B., {Phillips}, M.~M., {Reiss}, D.,
  {Schmidt}, B.~P., {Schommer}, R.~A., {Smith}, R.~C., {Spyromilio}, J.,
  {Stubbs}, C., {Suntzeff}, N.~B., \& {Tonry}, J. 1998, \aj, 116, 1009


\bibitem[{{Riess} {et~al.}(2004){Riess}, {Strolger}, {Tonry}, {Casertano},
  {Ferguson}, {Mobasher}, {Challis}, {Filippenko}, {Jha}, {Li}, {Chornock},
  {Kirshner}, {Leibundgut}, {Dickinson}, {Livio}, {Giavalisco}, {Steidel},
  {Ben{\'{\i}}tez}, \& {Tsvetanov}}]{Riess04}
{Riess}, A.~G., {Strolger}, L.-G., {Tonry}, J., {Casertano}, S., {Ferguson},
  H.~C., {Mobasher}, B., {Challis}, P., {Filippenko}, A.~V., {Jha}, S., {Li},
  W., {Chornock}, R., {Kirshner}, R.~P., {Leibundgut}, B., {Dickinson}, M.,
  {Livio}, M., {Giavalisco}, M., {Steidel}, C.~C., {Ben{\'{\i}}tez}, T., \&
  {Tsvetanov}, Z. 2004, \apj, 607, 665


\bibitem[{{Sako} {et~al.}(2008){Sako}, {Bassett}, {Becker}, {Cinabro},
  {DeJongh}, {Depoy}, {Dilday}, {Doi}, {Frieman}, {Garnavich}, {Hogan},
  {Holtzman}, {Jha}, {Kessler}, {Konishi}, {Lampeitl}, {Marriner}, {Miknaitis},
  {Nichol}, {Prieto}, {Riess}, {Richmond}, {Romani}, {Schneider}, {Smith},
  {Subba Rao}, {Takanashi}, {Tokita}, {van der Heyden}, {Yasuda}, {Zheng},
  {Barentine}, {Brewington}, {Choi}, {Dembicky}, {Harnavek}, {Ihara}, {Im},
  {Ketzeback}, {Kleinman}, {Krzesi{\'n}ski}, {Long}, {Malanushenko},
  {Malanushenko}, {McMillan}, {Morokuma}, {Nitta}, {Pan}, {Saurage}, \&
  {Snedden}}]{Sako08}
{Sako}, M., {Bassett}, B., {Becker}, A., {Cinabro}, D., {DeJongh}, F., {Depoy},
  D.~L., {Dilday}, B., {Doi}, M., {Frieman}, J.~A., {Garnavich}, P.~M.,
  {Hogan}, C.~J., {Holtzman}, J., {Jha}, S., {Kessler}, R., {Konishi}, K.,
  {Lampeitl}, H., {Marriner}, J., {Miknaitis}, G., {Nichol}, R.~C., {Prieto},
  J.~L., {Riess}, A.~G., {Richmond}, M.~W., {Romani}, R., {Schneider}, D.~P.,
  {Smith}, M., {Subba Rao}, M., {Takanashi}, N., {Tokita}, K., {van der
  Heyden}, K., {Yasuda}, N., {Zheng}, C., {Barentine}, J., {Brewington}, H.,
  {Choi}, C., {Dembicky}, J., {Harnavek}, M., {Ihara}, Y., {Im}, M.,
  {Ketzeback}, W., {Kleinman}, S.~J., {Krzesi{\'n}ski}, J., {Long}, D.~C.,
  {Malanushenko}, E., {Malanushenko}, V., {McMillan}, R.~J., {Morokuma}, T.,
  {Nitta}, A., {Pan}, K., {Saurage}, G., \& {Snedden}, S.~A. 2008, \aj, 135,
  348


\bibitem[{{Schlegel} {et~al.}(1998){Schlegel}, {Finkbeiner}, \&
  {Davis}}]{Schlegel98}
{Schlegel}, D.~J., {Finkbeiner}, D.~P., \& {Davis}, M. 1998, \apj, 500, 525


\bibitem[{{Sisson} {et~al.}(2007){Sisson}, {Fan}, \& {Tanaka}}]{Sisson07}
{Sisson}, S.~A., {Fan}, Y., \& {Tanaka}, M.~M. 2007, Proceedings of the
  National Academy of Science, 104, 1760


\bibitem[{{Stritzinger} {et~al.}(2011){Stritzinger}, {Phillips}, {Boldt},
  {Burns}, {Campillay}, {Contreras}, {Gonzalez}, {Folatelli}, {Morrell},
  {Krzeminski}, {Roth}, {Salgado}, {DePoy}, {Hamuy}, {Freedman}, {Madore},
  {Marshall}, {Persson}, {Rheault}, {Suntzeff}, {Villanueva}, {Li}, \&
  {Filippenko}}]{Stritzinger11}
{Stritzinger}, M.~D., {Phillips}, M.~M., {Boldt}, L.~N., {Burns}, C.,
  {Campillay}, A., {Contreras}, C., {Gonzalez}, S., {Folatelli}, G., {Morrell},
  N., {Krzeminski}, W., {Roth}, M., {Salgado}, F., {DePoy}, D.~L., {Hamuy}, M.,
  {Freedman}, W.~L., {Madore}, B.~F., {Marshall}, J.~L., {Persson}, S.~E.,
  {Rheault}, J.-P., {Suntzeff}, N.~B., {Villanueva}, S., {Li}, W., \&
  {Filippenko}, A.~V. 2011, \aj, 142, 156


\bibitem[{Tavare {et~al.}(1997)Tavare, Balding, Griffiths, \&
  Donnelly}]{Tavare97}
Tavare, S.~A., Balding, D.~J., Griffiths, R.~C., \& Donnelly, P. 1997,
  Genetics, 145, 505


\bibitem[{Wasserman(2006)}]{Wasserman06}
Wasserman, L. 2006, All of Nonparametric Statistics (233 Spring Street, New
  York, NY 10013,USA: Springer Science+Businesse Media, Inc.)


\bibitem[{Weiss \& von Haeseler(1998)}]{Weiss98}
Weiss, G. \& von Haeseler, A. 1998, Genetics, 149, 1539
 \href{http://www.genetics.org/content/149/3/1539.abstract}{\urllinklabel}

\bibitem[{{Wood-Vasey} {et~al.}(2007){Wood-Vasey}, {Miknaitis}, {Stubbs},
  {Jha}, {Riess}, {Garnavich}, {Kirshner}, {Aguilera}, {Becker}, {Blackman},
  {Blondin}, {Challis}, {Clocchiatti}, {Conley}, {Covarrubias}, {Davis},
  {Filippenko}, {Foley}, {Garg}, {Hicken}, {Krisciunas}, {Leibundgut}, {Li},
  {Matheson}, {Miceli}, {Narayan}, {Pignata}, {Prieto}, {Rest}, {Salvo},
  {Schmidt}, {Smith}, {Sollerman}, {Spyromilio}, {Tonry}, {Suntzeff}, \&
  {Zenteno}}]{Wood-Vasey07}
{Wood-Vasey}, W.~M., {Miknaitis}, G., {Stubbs}, C.~W., {Jha}, S., {Riess},
  A.~G., {Garnavich}, P.~M., {Kirshner}, R.~P., {Aguilera}, C., {Becker},
  A.~C., {Blackman}, J.~W., {Blondin}, S., {Challis}, P., {Clocchiatti}, A.,
  {Conley}, A., {Covarrubias}, R., {Davis}, T.~M., {Filippenko}, A.~V.,
  {Foley}, R.~J., {Garg}, A., {Hicken}, M., {Krisciunas}, K., {Leibundgut}, B.,
  {Li}, W., {Matheson}, T., {Miceli}, A., {Narayan}, G., {Pignata}, G.,
  {Prieto}, J.~L., {Rest}, A., {Salvo}, M.~E., {Schmidt}, B.~P., {Smith},
  R.~C., {Sollerman}, J., {Spyromilio}, J., {Tonry}, J.~L., {Suntzeff}, N.~B.,
  \& {Zenteno}, A. 2007, \apj, 666, 694


\end{thebibliography}

\end{document}